\documentstyle[apjfonts,emulateapj]{article}

\lefthead{KASPI ET AL.}

\righthead{REVERBERATION AND SIZE--MASS--LUMINOSITY}

\slugcomment{Accepted Version (November 22, 1999)}

\def\gtorder{\mathrel{\raise.3ex\hbox{$>$}\mkern-14mu
                \lower0.6ex\hbox{$\sim$}}}
\def\ltorder{\mathrel{\raise.3ex\hbox{$<$}\mkern-14mu
                \lower0.6ex\hbox{$\sim$}}}

\def\Ha{H$\alpha$}
\def\Hb{H$\beta$}
\def\Hg{H$\gamma$}
\def\ergscmA{ergs\,s$^{-1}$\,cm$^{-2}$\,\AA$^{-1}$}
\def\ergscm{ergs\,s$^{-1}$\,cm$^{-2}$}
\def\Rblr{$R_{BLR}$}
\def\ha{H$\alpha$}

\begin{document}

\title{Reverberation Measurements for 17 Quasars and 
the Size--Mass--Luminosity Relations in Active Galactic Nuclei}
\author{
Shai~Kaspi,\altaffilmark{1,2}
Paul~S.~Smith,\altaffilmark{3}
Hagai~Netzer,\altaffilmark{1}
Dan~Maoz,\altaffilmark{1}
Buell~T.~Jannuzi,\altaffilmark{3}
and Uriel~Giveon\altaffilmark{1}}
\altaffiltext{1}{School of Physics and Astronomy and the Wise
Observatory, The Raymond and Beverly Sackler Faculty of Exact Sciences,
Tel-Aviv University, Tel-Aviv 69978, Israel.}
\altaffiltext{2}{Current address: Department of Astronomy \& Astrophysics,
The Pennsylvania State University, University Park, PA 16802. E-mail: shai@astro.psu.edu.}
\altaffiltext{3}{National Optical Astronomy Observatories, P.O. Box 26732, Tucson, AZ 85719.}
%
% \authoremail{shai@wise.tau.ac.il}

\begin{abstract}

Correlated variations in the line and continuum emission from active
galactic nuclei (AGN) can be used to determine the size and geometry of
the broad emission line regions (BLRs).  We have spectrophotometrically
monitored a well-defined sample of 28 Palomar-Green quasars in order to
obtain measurements of their BLRs and to investigate the relationships
between quasar luminosity, central black hole mass, and BLR size in
AGN. Spectrophotometry was obtained every 1--4 months for 7.5 years,
yielding 20--70 observing epochs per object.  Both the continuum and
emission line fluxes of all of the quasars were observed to change
during the duration of the observing program. Seventeen of the 28
objects were observed with adequate sampling ($\gtorder 20$ independent
observing epochs) to search for correlated variations between the
Balmer emission lines and the continuum flux. For each of these 17
objects, a significant correlation was observed, with the Balmer line
variations lagging those of the continuum by $\sim$100 days (rest
frame).  Our work increases the available luminosity range for studying
the size--mass--luminosity relations in AGN by two orders of magnitude
and doubles the number of objects suitable for such studies. Combining
our results with comparable published data available for Seyfert 1
galaxies, we find the BLR size scales with the rest-frame
5100~\AA\ luminosity as $L^{0.70\pm 0.03}$. This determination of the
scaling of the size of the BLR as a function of luminosity is
significantly different from those previously published, and suggests
that the effective ionization parameter in AGN may be a decreasing
function of luminosity. We are also able to constrain, subject to our
assumption that gravity dominates the motions of the BLR gas, the
scaling relationship between the mass of the central black holes and
the luminosity in AGN. We find that the central mass scales with 5100
\AA\ luminosity as $M\propto L^{0.5 \pm 0.1}$. This is inconsistent
with all AGN having optical luminosity that is a constant fraction of
the Eddington luminosity.
\end{abstract}


\section{Introduction}

Reverberation mapping, observing the degree and nature of the
correlation between continuum and emission line flux variations, is one
of the major tools for studying the distribution and kinematics of the
gas in the broad line region (BLR) of active galactic nuclei (AGN; see
reviews by Peterson 1993 and by Netzer \& Peterson 1997). About 17 low
luminosity AGN (Seyfert~1 galaxies) have been successfully monitored
and produce statistically meaningful results (see Wandel, Peterson, \&
Malkan 1999, and references therein). Best studied among these is the
Seyfert 1 galaxy, NGC~5548, which was monitored from the ground for
over eight years, and from space for several long periods (Peterson et
al. 1999, and references therein). Several other Seyfert~1s were
observed for periods of order 1 year or less, and nine Seyfert 1s were
studied over a period of eight years (Peterson et al. 1998a). The
measured time lags between the emission lines and the continuum light
curves in these objects can be interpreted in terms of the delayed
response of a spatially-extended BLR to a variable, compact source of
ionizing radiation. While the observations do not uniquely determine
the geometry of the BLR, they give its typical size which, for
Seyfert~1 galaxies, is  of the order of light-days to several
light-weeks ($\sim 10^{16}$--$10^{17}$ cm).  Recent studies have shown
that the time lags determined in NGC 5548 for different observing
seasons correlate with the seasonal luminosity of the object (Peterson
et al. 1999), and have presented evidence for Keplerian motions of the
BLR gas (Peterson \& Wandel 1999).

While Seyfert~1 galaxies have been studied successfully using
reverberation mapping, few similar studies of the more luminous AGN --
the quasars -- have been presented.  Those that have been published
have been limited in their success at determining the properties of the
quasar BLRs. Past attempts to monitor quasars spectrophotometrically
have generally suffered from temporal sampling and/or flux calibrations
that are not sufficient for the determination of the BLR size.  Zheng
and collaborators (Zheng \& Burbidge 1986, Zheng et al. 1987, and Zheng
1988) reported results of a monitoring program executed on 30 quasars
for several years, with a sampling interval of about one year.  They
found the emission-line flux to change in response to the continuum
changes, but because of the limited time resolution, only an upper
limit of about one light-year for the BLR size could be deduced.
Several other groups (P\'{e}rez, Penston, \& Moles 1989, Korista 1991,
Jackson et al. 1992, and Erkens et al. 1995) observed samples of
quasars for approximately one year sampled with monthly observations.
Each of these groups reported line variations on time scales of a few
months, but because of the short duration of these programs the data
were insufficient for more specific conclusions. Wisotzki et~al. (1998)
monitored the gravitationally lensed quasar HE\,1104$-$1805 for five
years. Although the quasar continuum varied considerably, the emission
line fluxes appeared to remain constant.  Researchers have also
attempted to use spectra obtained with the {\em International
Ultraviolet Explorer (IUE)} and reverberation mapping techniques to
study AGN. In particular, several studies (Gondhalekar et al. 1986,
Gondhalekar 1990, Sitko 1990, O'Brien \& Gondhalekar 1991, and Koratkar
et al. 1998) deduced BLR sizes of 0.01--2 light-years for AGN. However,
these results are controversial due to insufficient sampling and/or
uncertainty regarding the absolute flux calibrations of the IUE
observations (e.g., Bohlin \& Grillmair 1988a, 1988b).  Even the quasar
best studied by IUE, 3C\,273, has yielded disputed results when
different researchers have analyzed similar IUE data sets. Both O'Brien
\& Harries (1991) and Koratkar \& Gaskell (1991a) found a measurable
and similar lag between continuum and BLR variations, while Ulrich,
Courvoisier, \& Wamsteker (1993) argue that the line variations
reported in the earlier studies were only marginally significant. Thus,
even in the the best-studied case, past results have proven
controversial.

To obtain more definite results on the BLR size in quasars, we began
monitoring a well-defined sub-sample of 28 quasars from the
Palomar-Green (PG) sample (Schmidt \& Green 1983) in 1991 March at the
Wise and Steward Observatories.  Results from the first 1.5 years were
presented in Maoz et al. (1994; Paper~I) where it was shown that most
quasars  underwent continuum variations (at 4800 \AA\ rest wavelength)
with amplitudes of 10\%~--~70\%. Balmer-line variability, correlated
with the continuum variations, was  detected in several objects. The
preliminary data showed that the typical response time of the emission
lines is $\ltorder$\,6 months. Reverberation mapping of such objects
therefore requires several years, with sampling intervals of less than
a few months. Further results, based on four years of monitoring two
quasars from our sample, PG\,0804+761 and PG\,0953+414, were presented
in Kaspi et al. (1996a; Paper~II). The measured time lags between the
Balmer lines and the continuum variations of the quasars suggested that
the BLR size grows roughly as the square root of the luminosity of the
nucleus.  Wandel et al. (1999) confirmed this result using a larger
sample of AGN.

During the period of our spectrophotometric project we also monitored
several quasar samples photometrically in the $B$ and $R$ bands.  The
broad-band results for radio- and optically-selected samples were
presented in Netzer et al. (1996) and Giveon et al. (1999; hereafter
Paper III), respectively. The optical sample in Paper III consists of
42 PG quasars, including all 28 objects discussed in the present
paper.

The present paper presents final results from 7.5 years of
spectroscopic monitoring of our sample. In \S~2 we describe the sample,
observations, data reduction, and present the light curves. In \S~3 we
perform a time series analysis to determine the response time of the
emission-line flux to continuum variations. We estimate the BLR size,
continuum luminosity, and central mass for each quasar in \S~4, and in
\S~5 we discuss the relations among these properties.

\section{Sample, Observations, and Reduction}
\label{secobsred}

The sample, observing technique, and the reduction procedure, are
described in detail in Papers~I,~II,~\&~III. For the sake of
completeness, we briefly repeat them here, with emphasis on the few
procedures which, in the final reduction, were modified with respect to
previous descriptions.  The sample consists of 28 PG quasars.  Objects
with a northern declination, $B<16$ mag (based on the magnitudes given
in Schmidt \& Green 1983), redshift $z<0.4$, and a bright comparison
star within $3\farcm5$ of the quasar, were selected.
Table~\ref{tsample} presents detailed information about the sample.
Throughout this paper we will use only the first four digits in the PG
object name to identify individual objects.

\footnotesize
%\scriptsize
\begin{deluxetable}{cccccccccccc}
\tablecolumns{12}
\tablewidth{0pt}
\tablecaption{Sample Properties
\label{tsample}}
\tablehead{
\colhead{Object} 	&
\multicolumn{2}{c}{RA \ \ (1950.0) \ \ Dec} &
\colhead{z} 		&
\colhead{$m_B$}		&
\colhead{$M_B$}		&
\colhead{$A_B$}		&
\colhead{$N_{phot}$}	&
\colhead{$N_{spec}$}	&
\colhead{$f_{\lambda}$(5100\AA )} &
\colhead{$R_{comp}$} 	&
\colhead{PA$_{comp}$} 	\nl
\colhead{(1)} & 
\colhead{(2)} & 
\colhead{(3)} &
\colhead{(4)} & 
\colhead{(5)} & 
\colhead{(6)} & 
\colhead{(7)} &
\colhead{(8)} &
\colhead{(9)} &
\colhead{(10)} &
\colhead{(11)} &
\colhead{(12)} 
} 
\startdata
PG\,0026+129 & 00 26 38.1 & 12 59 30 & 0.142 & 15.3 & -23.7 & 0.13 & 38 & 56 & $ 26.9\pm  4.0$ &  95 &  42.0 \nl
PG\,0052+251 & 00 52 11.1 & 25 09 24 & 0.155 & 14.7 & -24.1 & 0.12 & 41 & 56 & $ 20.7\pm  3.7$ &  92 & 153.4 \nl
PG\,0804+761 & 08 04 35.4 & 76 11 32 & 0.100 & 14.5 & -23.7 & 0.11 & 38 & 70 & $ 54.8\pm 10.0$ & 182\tablenotemark{a} & 315.6\tablenotemark{a} \nl
PG\,0844+349 & 08 44 33.9 & 34 56 09 & 0.064 & 15.1 & -22.1 & 0.08 & 31 & 49 & $ 37.1\pm  3.8$ & 124 &  36.8 \nl
PG\,0953+414 & 09 53 48.3 & 41 29 58 & 0.239 & 15.4 & -24.5 & 0.00 & 38 & 36 & $ 15.6\pm  2.1$ & 104 &  31.7 \nl
PG\,1048+342 & 10 48 56.1 & 34 15 23 & 0.167 & 17.7 & -21.9 & 0.00 & 30 &  1 & $ 5.51\pm 0.28$ & 186 &  94.7 \nl
PG\,1100+772 & 11 00 27.4 & 77 15 09 & 0.313 & 15.9 & -24.7 & 0.06 & 46 &  9 & $ 7.72\pm 0.23$ &  68 &  84.1 \nl
PG\,1202+281 & 12 02 08.9 & 28 10 54 & 0.165 & 16.2 & -23.0 & 0.03 & 37 &  9 & $  7.2\pm  2.0$ &  73 & 219.8 \nl
PG\,1211+143 & 12 11 44.8 & 14 19 53 & 0.085 & 14.4 & -23.4 & 0.13 & 52 & 38 & $ 56.6\pm  9.2$ & 162 & 352.2 \nl
PG\,1226+023 & 12 26 33.4 & 02 19 42 & 0.158 & 12.8 & -26.3 & 0.00 & 26 & 39 & $  213\pm   26$ & 171 & 171.2 \nl
PG\,1229+204 & 12 29 33.1 & 20 26 03 & 0.064 & 15.5 & -21.7 & 0.00 & 32 & 33 & $ 21.5\pm  2.3$ &  75 & 291.5 \nl
PG\,1307+085 & 13 07 16.2 & 08 35 47 & 0.155 & 15.6 & -23.5 & 0.02 & 35 & 23 & $ 17.9\pm  1.8$ & 138 & 186.5 \nl
PG\,1309+355 & 13 09 58.5 & 35 31 15 & 0.184 & 15.5 & -24.0 & 0.00 & 30 &  9 & $ 23.6\pm  1.4$ & 167 &  74.2 \nl
PG\,1322+659 & 13 22 08.5 & 65 57 25 & 0.168 & 16.0 & -23.3 & 0.05 & 27 &  8 & $11.60\pm 0.74$ & 104 & 277.5 \nl
PG\,1351+640 & 13 51 46.3 & 64 00 28 & 0.087 & 14.7 & -23.2 & 0.03 & 40 & 30 & $ 51.4\pm  5.1$ & 109 & 196.5 \nl
PG\,1354+213 & 13 54 11.5 & 21 18 30 & 0.300 & 17.3 & -23.2 & 0.04 & 25 &  3 & $11.14\pm 0.25$ & 125 & 133.4 \nl
PG\,1402+261 & 14 02 59.2 & 26 09 51 & 0.164 & 15.6 & -23.6 & 0.00 & 27 &  6 & $18.73\pm 0.87$ & 171 & 334.8 \nl
PG\,1404+226 & 14 04 02.5 & 22 38 03 & 0.098 & 16.5 & -21.8 & 0.01 & 27 &  8 & $11.40\pm 0.91$ &  77 & 223.0 \nl
PG\,1411+442 & 14 11 50.1 & 44 14 12 & 0.089 & 15.0 & -22.9 & 0.00 & 34 & 24 & $ 37.1\pm  3.2$ & 128 & 347.0 \nl
PG\,1415+451 & 14 15 04.3 & 45 09 57 & 0.114 & 16.3 & -22.3 & 0.00 & 29 &  3 & $ 15.2\pm  1.8$ & 189 &  95.2 \nl
PG\,1426+015 & 14 26 33.8 & 01 30 27 & 0.086 & 15.7 & -22.2 & 0.12 & 33 & 20 & $ 46.2\pm  7.1$ & 106 & 341.4 \nl
PG\,1444+407 & 14 44 50.2 & 40 47 37 & 0.267 & 17.4 & -23.0 & 0.00 & 27 &  5 & $ 9.98\pm 0.39$ & 158 & 155.4 \nl
PG\,1512+370 & 15 12 46.9 & 37 01 56 & 0.371 & 16.4 & -24.5 & 0.05 & 35 & 10 & $ 4.82\pm 0.76$ &  63 & 284.9 \nl
PG\,1613+658 & 16 13 36.3 & 65 50 38 & 0.129 & 14.9 & -23.8 & 0.04 & 42 & 48 & $ 34.9\pm  4.3$ & 101 & 164.2 \nl
PG\,1617+175 & 16 17 56.2 & 17 31 34 & 0.114 & 15.5 & -22.8 & 0.15 & 39 & 35 & $ 14.4\pm  2.5$ & 147 & 253.0 \nl
PG\,1700+518 & 17 00 13.4 & 51 53 37 & 0.292 & 15.4 & -25.2 & 0.02 & 41 & 39 & $ 22.0\pm  1.5$ & 104 & 183.5 \nl
PG\,1704+608 & 17 04 03.5 & 60 48 31 & 0.371 & 15.6 & -25.5 & 0.00 & 39 & 25 & $ 16.6\pm  2.4$ & 114 & 188.2 \nl
PG\,2130+099 & 21 30 01.2 & 09 55 01 & 0.061 & 14.7 & -22.4 & 0.17 & 47 & 64 & $ 48.4\pm  4.5$ &  86 &  68.2 \nl
\enddata
\footnotesize
%\scriptsize
\tablenotetext{}{Notes.--- \\
(1) Object name, as it appears in Schmidt \& Green (1983). \\
(2)--(3) Right Ascension and Declination (1950). \\
(4) Redshift. \\
(5)--(6) Median (over all epochs) apparent and absolute $B$ magnitudes
from Paper~III. \\
(7) Galactic $B$-band extinction, $A_{B}$, from NED. See \S~\ref{seclum}. \\
(8)--(9) Number of photometric and spectrophotometric observing epochs.  \\
(10) Average flux density and its rms 
at rest wavelength $\sim$5100 \AA \ 
in units of $10^{-16}$ \ergscmA (see Table~\ref{tabwave} column (7)
for exact wavelength intervals) . \\
(11)--(12) The angular separation in arcseconds and position angle in
degrees of the comparison star (see \S~\ref{secobsred}).}
\tablenotetext{a}{Comparison star at SO: $R_{comp}=119^{\prime\prime}$,
PA$_{comp}=64.3$ degrees.}
\end{deluxetable}

%\footnotesize
\scriptsize
\begin{deluxetable}{cccccccccc}
\tablecolumns{10}
\tablewidth{0pt}
\tablecaption{\vglue -0.1cm
Integration Limits for Continuum Bands and Emission Lines
\label{tabwave}}
\tablehead{
\colhead{Object}    &
\colhead{Continuum} & 
\colhead{\Hg}       & 
\colhead{Continuum} &
\colhead{Continuum} & 
\colhead{\Hb}       & 
\colhead{Continuum} &
\colhead{Continuum} & 
\colhead{\Ha}       & 
\colhead{Continuum} \nl
\colhead{(1)} & 
\colhead{(2)} & 
\colhead{(3)} & 
\colhead{(4)} & 
\colhead{(5)} & 
\colhead{(6)} & 
\colhead{(7)} & 
\colhead{(8)} & 
\colhead{(9)} & 
\colhead{(10)} 
} 
\startdata
PG\,0026 & 4790--4830 & 4920--5045 & 5070--5100 & {\bf 5408--5468} & 5508--5654 & 5818--5868 & 6998--7138 & 7304--7728 & 7730--7790 \nl
PG\,0052 & {\bf 4864--4914} & 4926--5118 & 5418--5488 & 5418--5488 & 5518--5698 & 5908--5978 & 7028--7108 & 7398--7724 & 7725--7800 \nl
PG\,0804 & 4680--4705 & 4710--4850 & 4880--4905 & {\bf 5224--5264} & 5266--5474 & 5598--5630 & 6746--6814 & 6940--7412 & 7472--7520 \nl
PG\,0844 & 4503--4535 & 4536--4690 & {\bf 4710--4750} & 5048--5078 & 5098--5248 & 5414--5440 & 6708--6764 & 6848--7120 & 7248--7348 \nl
PG\,0953 & {\bf 5238--5278} & 5284--5450 & 5472--5506 & 5846--5892 & 5906--6086 & 6288--6320 & \nodata    & \nodata    & \nodata    \nl
PG\,1211 & 4525--4560 & 4625--4755 & 4780--4810 & {\bf 5120--5160} & 5180--5340 & 5490--5530 & 6670--6720 & 6900--7310 & 7480--7540 \nl
PG\,1226 & 4925--4945 & 4946--5130 & {\bf 5135--5165} & 5494--5522 & 5530--5756 & 5888--5928 & 7034--7085 & 7250--7788 & 7789--7808 \nl
PG\,1229 & 4500--4560 & 4560--4690 & 4694--4758 & 5052--5088 & 5094--5244 & {\bf 5412--5456} & 6562--6654 & 6838--7148 & 7244--7320 \nl
PG\,1307 & 4870--4915 & 4938--5088 & 5104--5154 & {\bf 5458--5488} & 5508--5698 & 5864--5914 & 7050--7100 & 7400--7740 & 7750--7800 \nl
PG\,1351 & 4585--4645 & 4646--4781 & 4815--4850 & {\bf 5048--5128} & 5196--5352 & 5545--5605 & 6638--6770 & 6898--7322 & 7362--7432 \nl
PG\,1411 & 4570--4600 & 4650--4790 & 4805--4835 & {\bf 5155--5185} & 5201--5382 & 5520--5555 & 6700--6760 & 6870--7360 & 7490--7560 \nl
PG\,1426 & 4565--4605 & 4605--4775 & 4730--4796 & {\bf 5105--5145} & 5170--5360 & 5515--5555 & 6680--6730 & 6930--7270 & 7490--7550 \nl
PG\,1613 & \nodata    & \nodata    & \nodata    & {\bf 5300--5350} & 5360--5566 & 5740--5790 & 7000--7050 & 7013--7742 & 7743--7803 \nl
PG\,1617 & 4650--4700 & 4745--4950 & 4930--4960 & {\bf 5250--5295} & 5310--5550 & 5672--5722 & 6898--6968 & 7118--7498 & 7540--7580 \nl
PG\,1700 &  \nodata   & \nodata    & \nodata    & {\bf 6030--6080} & 6120--6410 & 6520--6570 & 7760--8015 & 8245--8689 & 8690--8730 \nl
PG\,1704 & 5800--5865 & 5900--6025 & {\bf 6040--6100} & 6460--6520 & 6575--6735 & 7020--7100 & \nodata    & \nodata    & \nodata    \nl
PG\,2130 & 4460--4490 & 4533--4710 & 4710--4735 & {\bf 5048--5084} & 5094--5240 & 5394--5440 & 6498--6598 & 6798--7118 & 7246--7318 \nl
\enddata
\footnotesize
\tablenotetext{}{Note.---Wavelengths in units of \AA \ at the observer's frame.
The bold-font ranges are the continuum bands shown in Fig.~\ref{flcs}.}
\end{deluxetable}
\normalsize

\begin{figure*}
\vglue -1in
\centerline{\epsfxsize=21cm\epsfbox{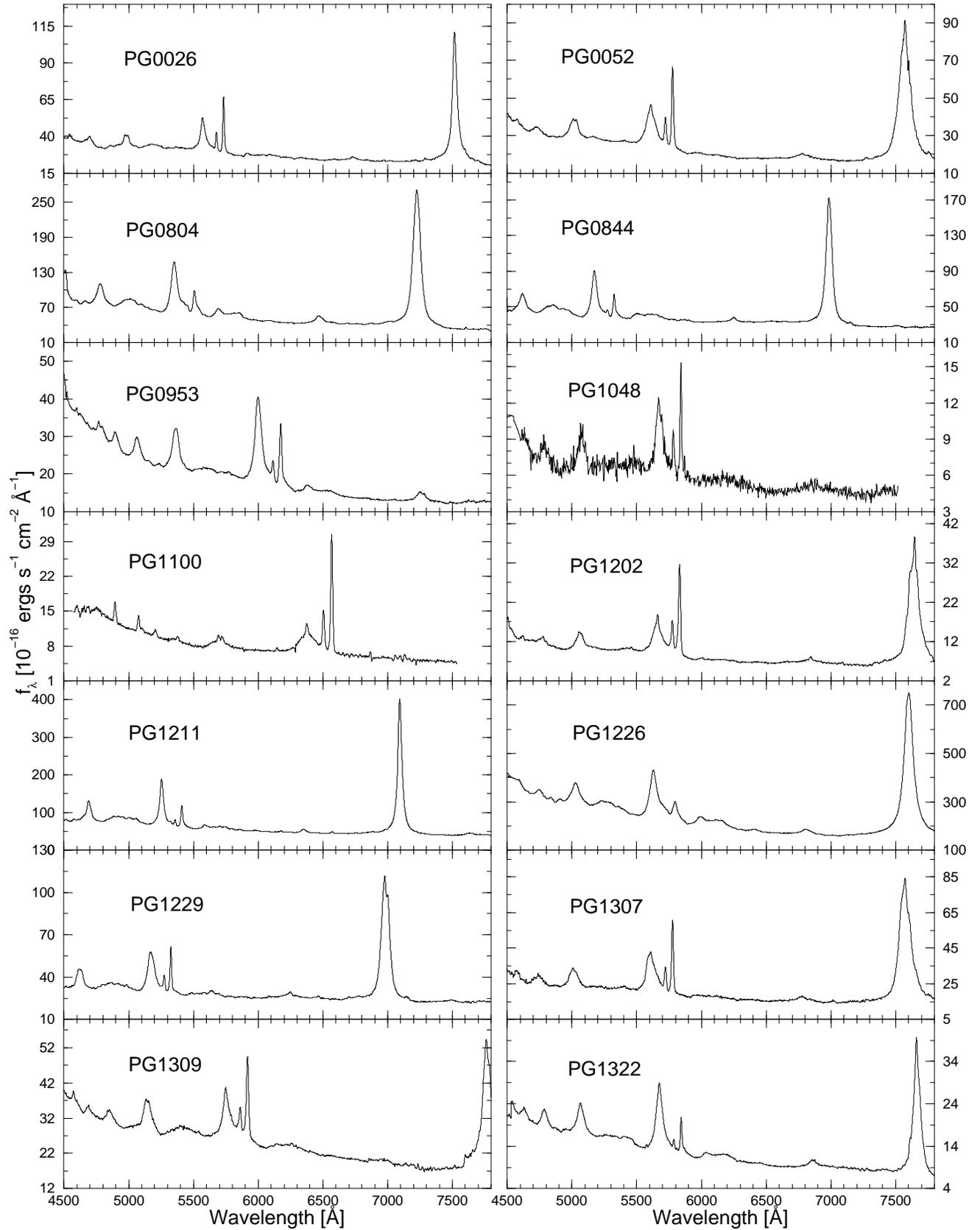}}
\vglue -0.7in
\caption{Mean spectra of PG quasars in the observed frame.}
\label{spectra1}
\end{figure*}

\addtocounter{figure}{-1}

\begin{figure*}
\vglue -1in
\centerline{\epsfxsize=21cm\epsfbox{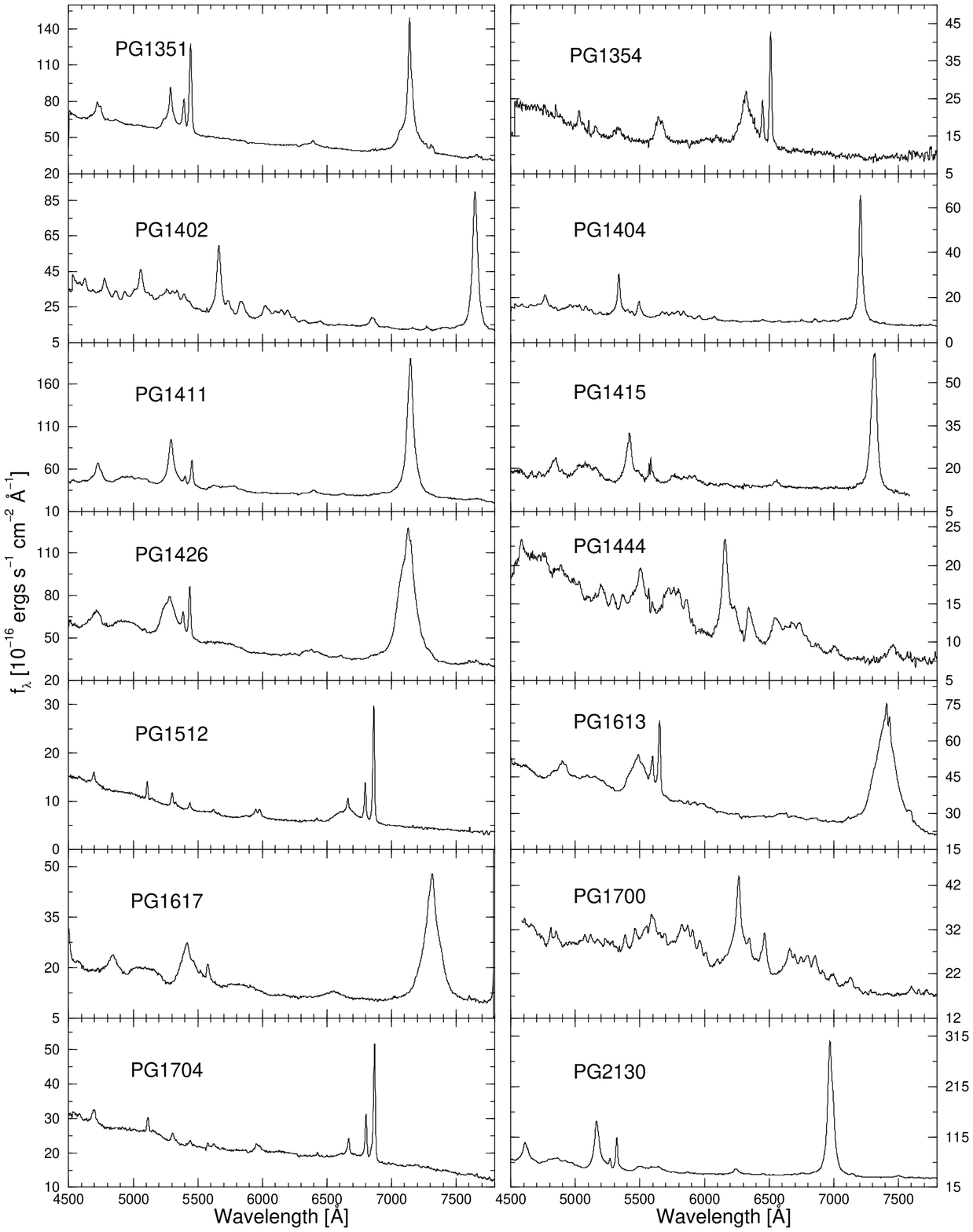}}
\vglue -0.7in
\caption{\it Continued}
\end{figure*}

The observations were carried out using the Steward Observatory (SO)
2.3m telescope and the Wise Observatory (WO) 1m telescope. Throughout
the years different detectors were used. At SO we used the Boller \&
Chivens spectrograph with a 300 l/mm grating, a $4\farcs5$-wide slit
and a TI CCD. In 1992 September this CCD was replaced with a
1200$\times$800 Loral CCD. At WO the Faint Object Spectroscopic Camera
(FOSC; Kaspi et al. 1995) was used with a 15$^{\prime\prime}$-wide slit
and a front-illuminated 1024$\times$1024 TI~CCD as a detector. In 1994
February the slit was replaced by a 10$^{\prime\prime}$-wide slit and
the detector used was a 1024$\times$1024 back-illuminated Tektronix
CCD. The typical wavelength coverage, at both observatories, was
$\sim$\,4000--8000~\AA , with a dispersion of $\sim$3.8 \AA/pixel, and
a spectral resolution of $\sim$\,10~\AA .

Spectrophotometric calibration for every quasar was accomplished by
rotating the spectrograph slit to the appropriate position angle so
that a nearby comparison star (columns [11] \& [12] in
Table~\ref{tsample}) was observed simultaneously with the object. A
wide slit was used to minimize the effects of atmospheric dispersion at
the non-parallactic position angle. This technique provides excellent
calibration even during poor weather conditions, and accuracies of
order 1\%~--~2\% can easily be achieved.

Observations typically consisted of two consecutive exposures of the
quasar/star pair. Total exposure times were usually 40 minutes at the
SO 2.3m telescope and 2 hours at the WO 1m telescope. The spectroscopic
data were reduced using standard IRAF\footnote{{IRAF (Image Reduction
and Analysis Facility) is distributed by the National Optical Astronomy
Observatories, which are operated by AURA, Inc., under cooperative
agreement with the National Science Foundation.}} routines. The
consecutive quasar/star flux ratios were compared to test for
systematic errors in the observations and to clean cosmic rays. The
ratio usually reproduced to 0.5--2.5\% at all wavelengths and
observations with ratios larger than 5\% were discarded. We verified
that none of our comparison stars are variable to within $\sim$2\% by
means of differential photometry with other stars in each field
(Paper~III), Spectra were calibrated to an absolute flux scale using
observations of spectrophotometric standard stars on one or more
epochs. The absolute flux calibration has an uncertainty of $\sim$10\%,
which is not shown in the error bars in our light curves. The error
bars reflect only the measurement and differential uncertainties, which
are of order 1\%--2\%.

\begin{figure*}[t]
\centerline{\epsfxsize=8.5cm\epsfbox{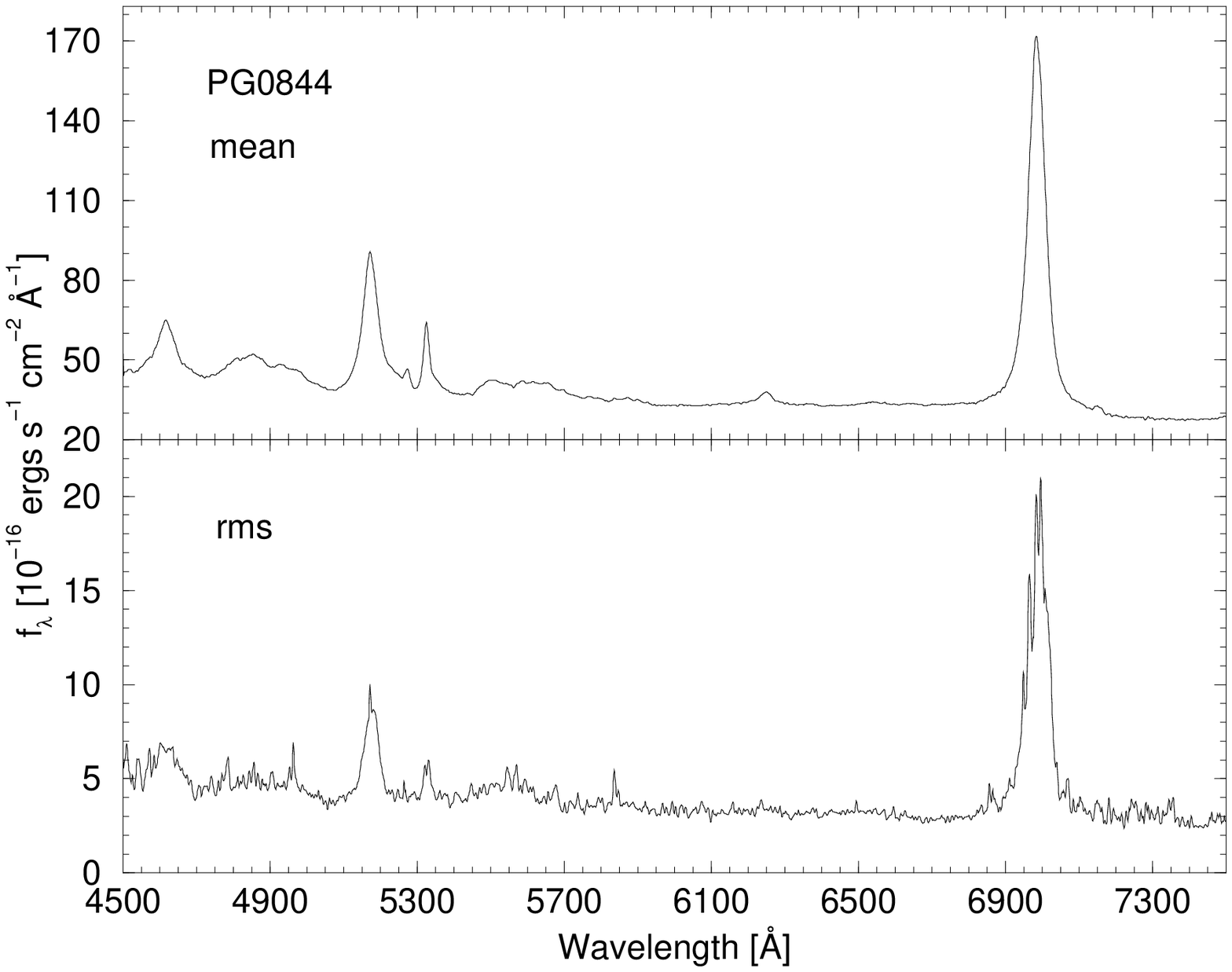}}
\caption{PG\,0844 mean spectrum ({\em top}) and rms spectrum
({\em bottom}).}
\label{pg0844spec}
\end{figure*}

For each quasar we used all the spectra to produce an average spectrum
and a root-mean-square (rms) spectrum, defined as:
\begin{equation}
\sigma(\lambda)=\left\{{1\over (N-1)}\sum_{i=1}^N \left[f_i(\lambda)-\bar 
f(\lambda)\right]^2\right\}^{1/2},
\end{equation}
where the sum is taken over the $N$ spectra, and $\bar f(\lambda)$ is
the average spectrum (Peterson et al. 1998a). The mean spectra of all
the quasars in our sample are shown in Fig.~\ref{spectra1}. An example
of a typical rms spectrum and comparison with the mean spectrum is
shown in Fig.~\ref{pg0844spec} for PG\,0844.  We used the average and
rms spectra to choose line-free spectral bands suitable for setting the
continuum underlying the emission lines, and the wavelength limits for
integrating the line fluxes. The spectral intervals for the Balmer
lines and the continuum bands on both sides of each line are given in
Table~\ref{tabwave}. The line and continuum fluxes were measured
algorithmically for all epochs by calculating the mean flux in the
continuum bands, and summing the flux above a straight line in
$f_{\lambda}$ that connects the continuum bands straddling the emission
line. This process measures the total emission in each line, i.e., the
flux of the broad component of the line together with its narrow
component.  The uncertainty in the line flux was estimated by
propagating the uncertainty in the setting of the continuum levels,
determined from the standard deviation of the mean in the continuum
bands. To this we added in quadrature our estimate for the uncertainty
in the differential spectrophotometry.

For the purpose of time series analysis, we generally chose a region
just blue of \Hb\ for the continuum band.  In several cases where the
blue side of \Hb\ has an inferior light curve (due to, e.g.,
atmospheric absorption bands and/or superposed emission line blends) we
chose continuum bands around \Hg. Table~\ref{tabwave} lists in bold
font the continuum band chosen for each object. Both the continuum
bands and the emission lines can be affected by the contribution of
variable Fe\,{\sc ii} lines.  However, correcting for this effect is
beyond the scope of the present study. Similarly, our Balmer line
measurements might be contaminated by narrow forbidden emission-lines.
Though we measure \Hb\ without the [O\,{\sc
iii}]$\,\lambda\lambda$4995, 5007 lines (since there is a clear
separation between them) we cannot do so for the [N\,{\sc
ii}]$\,\lambda\lambda$6548, 6583 lines blended with the \Ha\ line.
However, we note that contamination from narrow lines should not affect
our time series analysis since this analysis relies on the variable
parts of the lines, and the narrow lines have never been observed to
vary in AGN over relevant time scales (e.g., Kaspi et~al. 1996b).

The AGN continuum dominates the emission-line flux and host galaxy
starlight contribution to the broad-band photometry.  To improve the
sampling of the continuum light curves, we therefore merged the
photometric  $B$ band light curves of Paper III into the
spectrophotometric continuum light curves (see Paper~I).  The
photometric campaign lasted 7 years and we extended the observations of
the present 28 objects for six more months to better overlap with the
spectrophotometric campaign. Those last photometric observations were
reduced and added to the light curves as explained in Paper~III.

The photometric and spectrophotometric light curves of each object were
intercalibrated by comparing all pairs of spectrophotometric and
photometric observations separated in time by less than a few days
(typically $\sim$5 days, and in some cases up to $\sim$10 days). This
produced 10 to 15 pairs of points per object.  A linear least-squares
fit between the spectrophotometric continuum fluxes and the photometric
fluxes (the magnitudes of Paper~III translated to fluxes) was used to
merge the two light curves.

For each object, the total number of photometric and spectrophotometric
observations is given in Table~\ref{tsample}, columns (8) \& (9) (the
sum of the two columns gives the total number of points in the
continuum light curve). Note that for particular emission-lines in a
few objects, some data points are missing because of cosmic-rays
contaminations, insufficient wavelength coverage, or low S/N. Light
curves for the 17 objects that have more than 20 spectrophotometric
data points are presented in Fig.~\ref{flcs} with the data listed in
Table~\ref{contab} and Table~\ref{linetab}\,\footnote {A
machine-readable version of these data is available upon request from
the authors, or at  URL:  http://wise-obs.tau.ac.il/$\sim$shai/PG/}.
Each of the remaining 11 objects in our sample have less than 10
spectrophotometric observations over the 7.5 year period. Such rare
sampling is insufficient to determine the line-to-continuum lag and
these objects will not be discussed further in this paper.

\begin{figure*}
\vglue -1in
\centerline{\epsfxsize=21cm\epsfbox{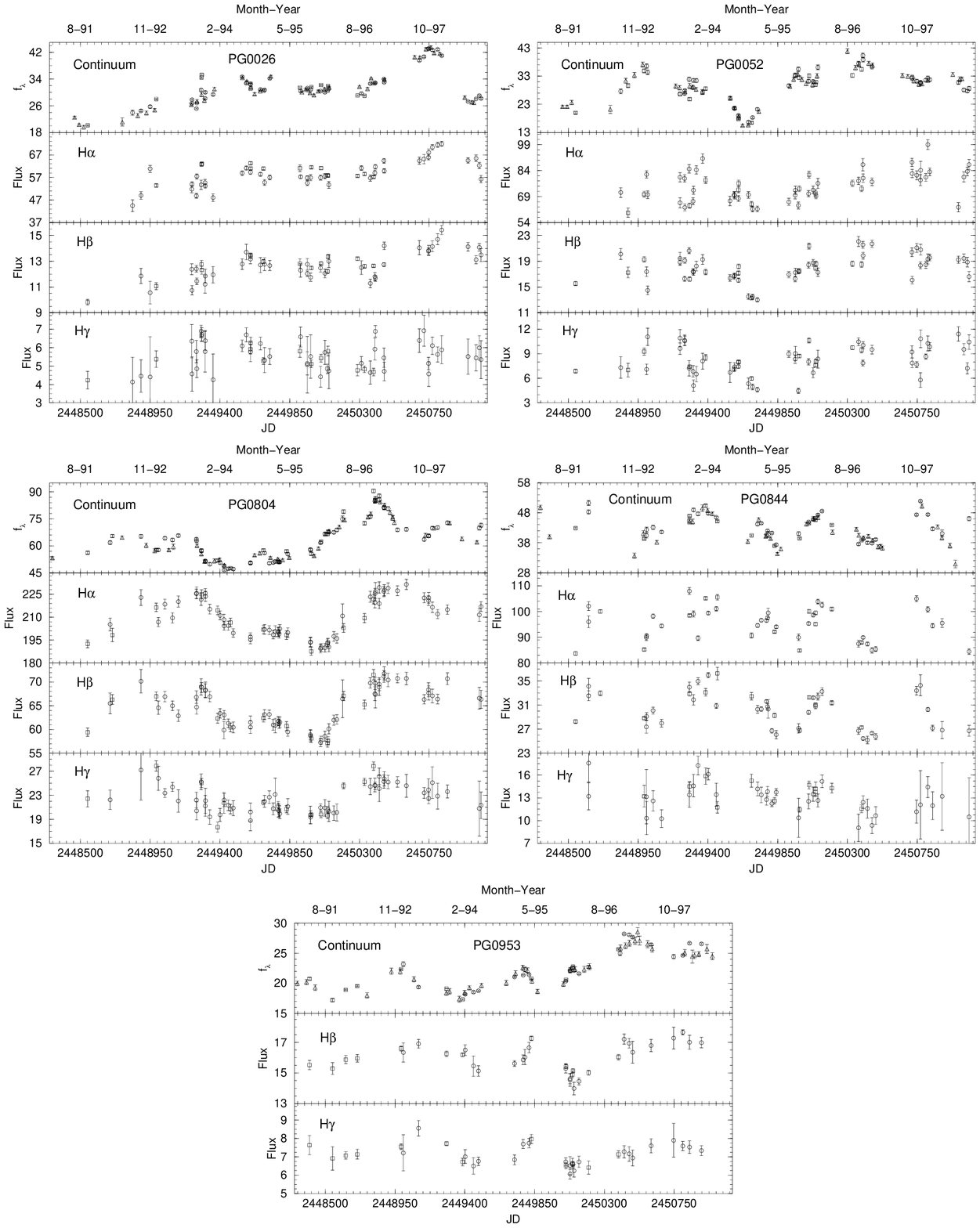}}
\vglue -0.7in
\caption{Light curves for 17 PG quasars. Circles are spectrophotometric
data from WO, squares are spectrophotometric data from SO, triangles
are photometric data from WO. Continuum flux densities,
f$_{\lambda}$, are determined from the wavelength bands listed in
bold-font in Table~\protect{\ref{tabwave}}, and are given in units of
10$^{-16}$\,\ergscmA . Emission-line fluxes are displayed in units of
10$^{-14}$\,\ergscm . Horizontal axis given in Julian Day (bottom) and
UT date (top).}
\label{flcs}
\end{figure*}

\addtocounter{figure}{-1}

\begin{figure*}
\vglue -1in
\centerline{\epsfxsize=21cm\epsfbox{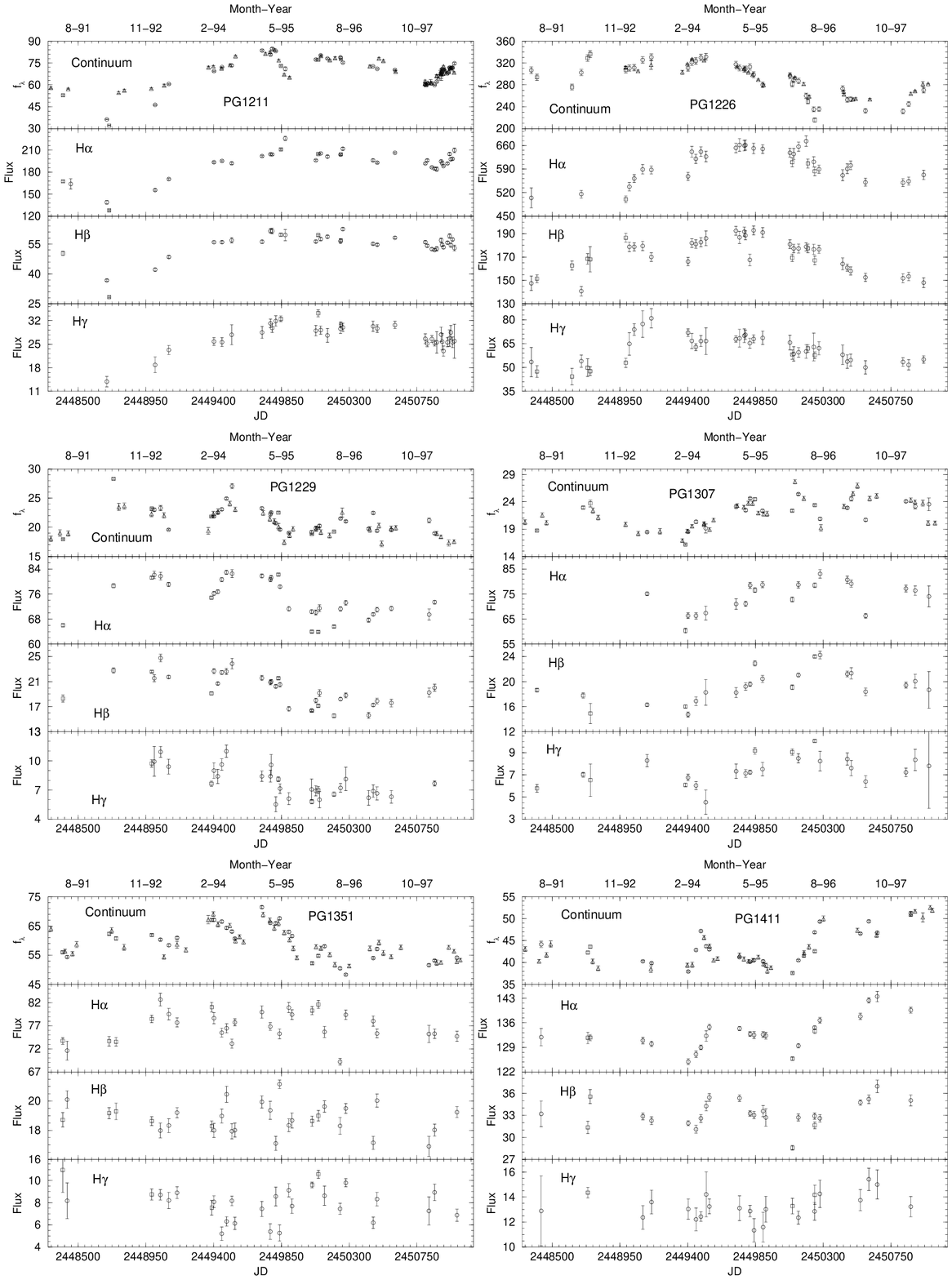}}
\vglue -0.5in
\caption{\it Continued}
\end{figure*}

\addtocounter{figure}{-1}

\begin{figure*}
\vglue -1in
\centerline{\epsfxsize=21cm\epsfbox{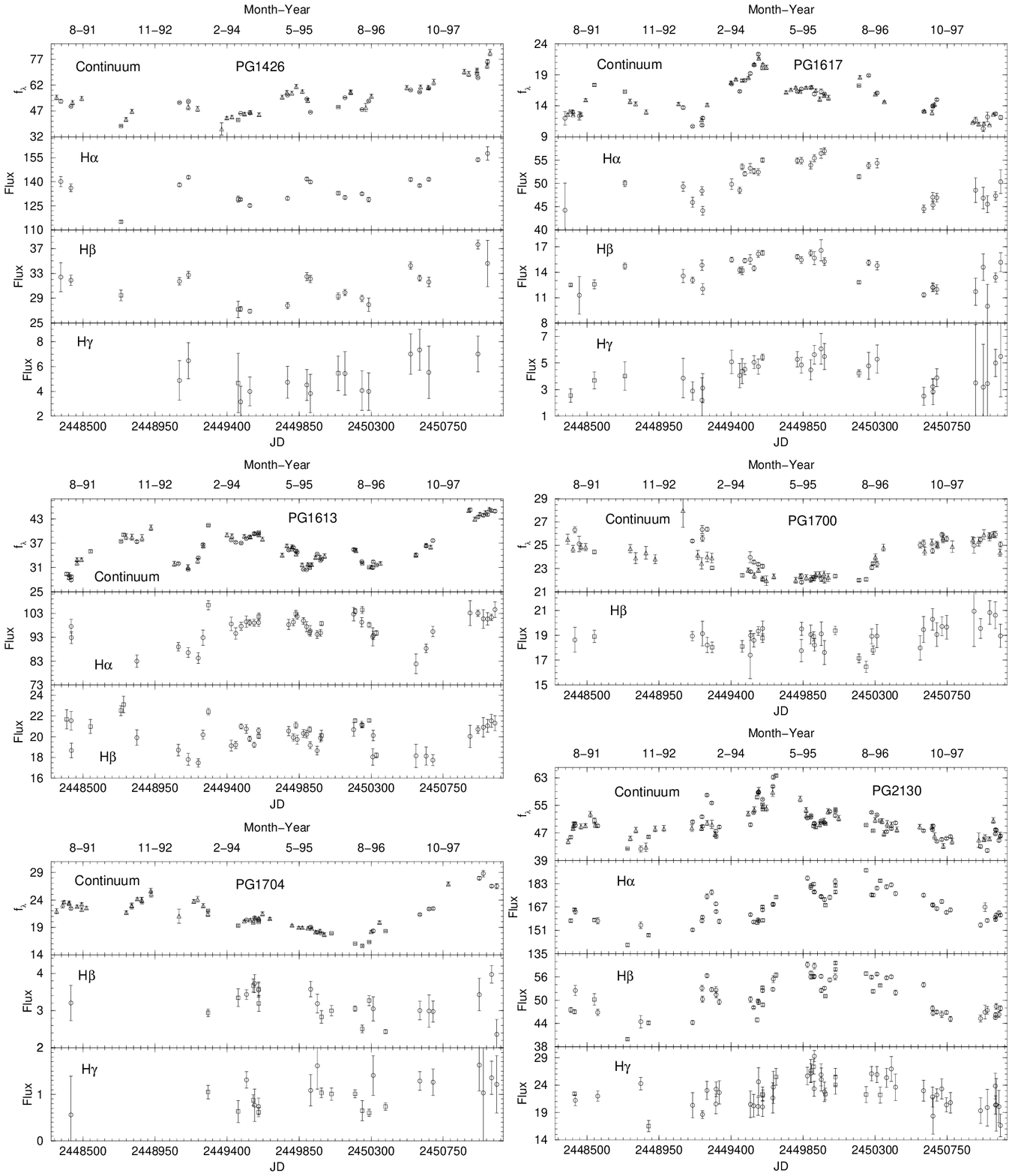}}
\vglue -0.7in
\caption{\it Continued}
\end{figure*}

\section{Time Series Analysis} 

The continuum variability properties of our quasar sample are
extensively discussed in Paper~III. All 17 objects underwent optical
continuum variability (at 5100 \AA\ rest wavelength) of 25\%--150\%,
measured relative to the minimum flux. In order to compare the line
variability to the continuum variability in each quasar, we define an
intrinsic normalized variability measure, $\sigma_{N} =
100\sqrt{\sigma^{2}-\delta^{2}}/\bar{f}$, where $\bar{f}$ and $\sigma$
are the average and the rms of the flux in a given light curve,
respectively, and $\delta$ is the mean uncertainty of all data points
in that light curve. This measure is less sensitive to noise and
outliers than the minimum-to-maximum flux ratio. The two cases where
${\sigma^{2}-\delta^{2}}$ is slightly negative (\Hg\ in PG\,1426 and
PG\,1617) are consistent with no variability. Table~\ref{variability}
summarizes $\sigma_{N}$ for the 5100 \AA\ rest frame continua band and
all of the available emission-line light curves. In general we find
that all of the Balmer emission lines in a specific object have about
the same level of variability. In all objects, the variability of the
emission lines is smaller than the continuum variability. In the few
cases where those trends do not hold, it appears that they may be
masked by noise (e.g., most cases in which the emission line varies
with a larger amplitude than the continuum involve \Hg , which has
lower S/N than \Ha\ and \Hb ).

The main goal of this project is to determine the time lag between
the continuum and the Balmer-line flux variations in high-luminosity
AGN. To quantify this lag, we use two methods for correlating the line
and continuum light curves. The first method is the interpolated
cross-correlation function (ICCF) of Gaskell \& Sparke (1986) and
Gaskell \& Peterson (1987), as implemented by White \& Peterson (1994;
see also a review by Gaskell 1994). The second method is the
$z$-transformed discrete correlation function (ZDCF) of Alexander
(1997) which is an improvement on the discrete correlation function
(DCF; Edelson \& Krolik 1988). The ZDCF applies Fisher's $z$
transformation to the correlation coefficients, and uses equal
population bins rather than the equal time bins used in the DCF. 
The two independent methods are in excellent agreement for
these data and we will use only the ICCF results in the following
analyses.

The uncertainties in the cross correlation lag determination can be
estimated using the model-independent FR/RSS Monte Carlo method
described by Peterson et al. (1998b). In this method, each Monte Carlo
simulation is composed of two parts: The first is a ``random subset
selection'' (RSS) procedure which consists of randomly drawing, with
replacement, from a light curve of $N$ points a new sample of $N$
points. After the $N$ points are selected, the redundant selections are
removed from the sample such that the temporal order of the remaining
points is preserved. This procedure reduces the number of points in
each light curve by a factor of $\sim 1/e$ and accounts for the effect
that individual data points have on the cross-correlation. The second
part is ``flux randomization'' (FR) in which the observed fluxes are
altered by random Gaussian deviates scaled to the uncertainty ascribed
to each point. This procedure simulates the effect of measurement
uncertainties.  Peterson et al. (1998b) demonstrate that under a wide
variety of realistic conditions, the combined FR/RSS procedure yields
conservative uncertainties compared to the 
true situation.

\scriptsize
%\footnotesize
\begin{deluxetable}{cc}
\tablecolumns{2}
\tablewidth{180pt}
\tablecaption{Continuum Light Curves
\label{contab}}
\tablehead{
\colhead{JD} 	&
\colhead{$f_{\lambda}$\tablenotemark{a}}
} 
\startdata
\multicolumn{2}{c}{PG\,0026}   \nl
2448461.5 & $ 22.48 \pm 0.36 $ \nl
2448490.6 & $ 20.30 \pm 0.30 $ \nl
2448520.3 & $ 19.59 \pm 0.56 $ \nl
2448545.8 & $ 20.11 \pm 0.15 $ \nl
2448769.6 & $ 21.12 \pm 1.20 $ \nl
2448836.5 & $ 23.88 \pm 0.74 $ \nl
2448869.4 & $ 23.02 \pm 0.64 $ \nl
2448890.4 & $ 24.51 \pm 0.31 $ \nl
2448925.2 & $ 23.83 \pm 0.58 $ \nl
2448949.4 & $ 25.73 \pm 0.46 $ \nl
\enddata
\footnotesize
\tablenotetext{a}{In units of 10$^{-16}$~ergs\,s$^{-1}$\,cm$^{-2}$\,\AA$^{-1}$.}
\tablenotetext{}{Note.---Table~\ref{contab} is presented in its
entirety in the electronic edition of the Astrophysical Journal. A
portion is shown here for guidance regarding its form and content.}
\end{deluxetable}

\begin{deluxetable}{cc}
\tablecolumns{2}
\tablewidth{180pt}
\tablecaption{Balmer Line Light Curves
\label{linetab}}
\tablehead{
\colhead{JD} 	&
\colhead{Flux\tablenotemark{a}}
} 
\startdata
\multicolumn{2}{c}{PG\,0026 - \ha}   \nl
2448836.5  &  $  44.50  \pm  2.56  $ \nl
2448890.4  &  $  49.02  \pm  1.60  $ \nl
2448949.4  &  $  60.81  \pm  1.56  $ \nl
2448988.6  &  $  53.46  \pm  0.71  $ \nl
2449218.6  &  $  51.81  \pm  1.57  $ \nl
2449220.5  &  $  54.04  \pm  0.99  $ \nl
2449248.4  &  $  48.81  \pm  1.03  $ \nl
2449250.5  &  $  57.43  \pm  1.14  $ \nl
2449278.6  &  $  53.72  \pm  2.00  $ \nl
2449281.7  &  $  62.68  \pm  0.70  $ \nl
\enddata
\footnotesize
\tablenotetext{a}{In units of 10$^{-14}$~ergs\,s$^{-1}$\,cm$^{-2}$.}
\tablenotetext{}{Note.---Table~\ref{linetab} is presented in its
entirety in the electronic edition of the Astrophysical Journal. A
portion is shown here for guidance regarding its form and content.}
\end{deluxetable}

\begin{deluxetable}{crrrr}
\tablecolumns{5}
\tablewidth{200pt}
\tablecaption{Variability Measure $\sigma_{N}$
\label{variability}}
\tablehead{
\colhead{Object} &
\colhead{Cont\tablenotemark{a}} &
\colhead{\Ha} &
\colhead{\Hb} &
\colhead{\Hg} % \nl
}
\startdata
PG\,0026 & 14.7  &  9.3    &  8.0    &  8.4    \nl
PG\,0052 & 17.6  & 10.1    & 11.4    & 20.2    \nl
PG\,0804 & 18.2  &  5.8    &  6.2    &  8.7    \nl
PG\,0844 & 10.1  &  6.7    &  9.7    & 10.0    \nl
PG\,0953 & 13.3  & \nodata &  5.5    &  6.2    \nl
PG\,1211 & 16.2  & 10.3    & 12.2    & 12.8    \nl
PG\,1226 & 11.9  &  7.7    &  7.8    & 12.3    \nl
PG\,1229 & 10.9  &  8.1    & 12.5    & 17.9    \nl
PG\,1307 &  9.8  &  7.9    & 12.8    & 14.3    \nl
PG\,1351 &  9.8  &  4.0    &  4.7    & \nodata \nl
PG\,1411 &  8.5  &  3.4    &  5.1    &  2.8    \nl
PG\,1426 & 15.3  &  7.0    &  8.8    & \nodata \nl
PG\,1613 & 12.4  &  5.6    &  6.1    & \nodata \nl
PG\,1617 & 17.3  &  7.4    & 11.5    & \nodata \nl
PG\,1700 &  6.8  & \nodata &  3.2    & \nodata \nl
PG\,1704 & 14.5  & \nodata & 10.7    & 18.0    \nl
PG\,2130 &  9.3  & 6.2     &  8.9    &  8.9    \nl
\enddata
%\small
\footnotesize
\tablenotetext{a}{Continuum at rest wavelength $\sim$5100~\AA \ (see
Table~\ref{tabwave} column [7] for exact wavelength intervals).}
\end{deluxetable}
\normalsize

%\scriptsize
\begin{deluxetable}{ccllccllc}
\tablecolumns{9}
\tablewidth{0pt}
\tablecaption{Cross-Correlation Results, Velocity Measurements, and Fluxes
\label{tccfs}}
\tablehead{
\colhead{Object} 	&
\colhead{line} 		&
\colhead{$\tau_{cent}$} &
\colhead{$\tau_{peak}$} &
\colhead{$r_{max}$}	&
\colhead{Number}	&
\colhead{$v_{FWHM}$(mean)}&
\colhead{$v_{FWHM}$(rms)} &
\colhead{Flux} 		\nl
\colhead{} 		&
\colhead{} 		&
\colhead{(days)} 	&
\colhead{(days)} 	&
\colhead{}		&
\colhead{Obs.}		&
\colhead{km\,s$^{-1}$}	&
\colhead{km\,s$^{-1}$}	&
\colhead{10$^{-14}$\ergscm} 	\nl
\colhead{(1)} & 
\colhead{(2)} & 
\colhead{(3)} &
\colhead{(4)} & 
\colhead{(5)} & 
\colhead{(6)} & 
\colhead{(7)} &
\colhead{(8)} & 
\colhead{(9)} 
} 
\startdata
PG\,0026 & \Ha & $132^{+ 29}_{ -31}$ & $108^{+ 63}_{ -47}$ & 0.81 & 55 & $1527\pm 99$ & $1179\pm  76$ & $ 59.0\pm  5.6$ \nl
PG\,0026 & \Hb & $125^{+ 29}_{ -36}$ & $ 85^{+ 36}_{ -28}$ & 0.84 & 53 & $2100\pm140$ & $1358\pm  91$ & $ 12.7\pm  1.1$ \nl
PG\,0052 & \Ha & $211^{+ 66}_{ -44}$ & $193^{+ 85}_{ -93}$ & 0.68 & 53 & $2690\pm510$ & $3280\pm 630$ & $ 74.6\pm  7.9$ \nl
PG\,0052 & \Hb & $ 99^{+ 30}_{ -31}$ & $ 75^{+ 22}_{ -15}$ & 0.75 & 56 & $3990\pm240$ & $4550\pm 270$ & $ 18.0\pm  2.1$  \nl
PG\,0052 & \Hg & $ 90^{+ 27}_{ -20}$ & $ 69^{+ 19}_{  -9}$ & 0.74 & 56 & $4050\pm300$ & $5120\pm 380$ & $  8.2\pm  1.8$  \nl
PG\,0804 & \Ha & $193^{+ 20}_{ -17}$ & $187^{+ 29}_{ -37}$ & 0.83 & 70 & $2757\pm 55$ & $2909\pm  58$ & $   209\pm 12$ \nl
PG\,0804 & \Hb & $151^{+ 26}_{ -24}$ & $118^{+ 34}_{ -30}$ & 0.85 & 70 & $2984\pm 51$ & $2430\pm  42$ & $ 64.5 \pm  4.2$  \nl
PG\,0804 & \Hg & $108^{+ 39}_{ -38}$ & $ 82^{+ 67}_{ -46}$ & 0.74 & 67 & $3120\pm150$ & $2470\pm 120$ & $ 22.4 \pm  2.4$ \nl
PG\,0844 & \Ha & $ 39^{+ 16}_{ -16}$ & $ 37^{+ 19}_{ -24}$ & 0.74 & 46 & $2210\pm100$ & $2700\pm 120$ & $ 95.3\pm  6.5$  \nl
PG\,0844 & \Hb & $ 13^{+ 14}_{ -11}$\,\tablenotemark{a} & $ 10^{+ 20}_{ -19}$ & 0.79 & 48 & $2730\pm120$ & $2830\pm 120$ & $ 30.2 \pm  3.0 $ \nl
PG\,0844 & \Hg & $ 31^{+ 77}_{ -49}$ & $ 28^{+ 89}_{ -43}$ & 0.58 & 45 & $3210\pm360$ & $4770\pm 530$ & $ 13.0 \pm  1.9 $ \nl
PG\,0953 & \Hb & $187^{+ 27}_{ -33}$ & $207^{+ 22}_{ -26}$ & 0.72 & 35 & $2885\pm 65$ & $2723\pm  62$ & $ 15.97\pm  0.93$ \nl
PG\,0953 & \Hg & $203^{+ 36}_{ -59}$ & $213^{+ 24}_{ -73}$ & 0.54 & 34 & $3110\pm130$ & $3860\pm 160$ & $  7.12\pm  0.56$ \nl
PG\,1211 & \Ha & $116^{+ 38}_{ -46}$ & $ 69^{+123}_{ -33}$ & 0.87 & 37 & $1479\pm 74$ & $1051\pm  53$ & $192\pm 20$ \nl
PG\,1211 & \Hb & $103^{+ 32}_{ -44}$ & $ 28^{+142}_{ -26}$ & 0.87 & 36 & $1832\pm 81$ & $1479\pm  66$ & $ 54.5\pm  6.7$  \nl
PG\,1211 & \Hg & $145^{+ 66}_{ -47}$ & $117^{+120}_{ -68}$ & 0.86 & 34 & $2380\pm150$ & $2200\pm 140$ & $ 27.3\pm  3.9$ \nl
PG\,1226 & \Ha & $514^{+ 65}_{ -64}$ & $447^{+191}_{ -65}$ & 0.71 & 34 & $2810\pm190$ & $1330\pm  88$ & $601\pm 48  $ \nl
PG\,1226 & \Hb & $382^{+117}_{ -96}$ & $342^{+145}_{ -89}$ & 0.78 & 39 & $3416\pm 72$ & $2742\pm  58$ & $172 \pm 14 $ \nl
PG\,1226 & \Hg & $307^{+ 57}_{ -86}$ & $306^{+ 52}_{ -61}$ & 0.79 & 39 & $3760\pm170$ & $3440\pm 160$ & $ 61.1\pm  8.7$ \nl
PG\,1229 & \Ha & $ 71^{+ 39}_{ -46}$ & $ 48^{+ 85}_{ -10}$ & 0.78 & 32 & $2960\pm110$ & $3350\pm 130$ & $ 74.7 \pm  6.1$ \nl
PG\,1229 & \Hb & $ 36^{+ 32}_{ -18}$ & $ 20^{+ 10}_{ -20}$ & 0.80 & 33 & $3440\pm120$ & $3490\pm 120$ & $ 19.8\pm  2.5 $  \nl
PG\,1229 & \Hg & $ 16^{+ 27}_{ -39}$ & $  7^{+ 36}_{ -49}$ & 0.81 & 29 & $3400\pm330$ & $3000\pm 290$ & $  7.8 \pm  1.5 $ \nl
PG\,1307 & \Ha & $179^{+ 94}_{-145}$ & $228^{+ 97}_{-169}$ & 0.67 & 20 & $3690\pm140$ & $3280\pm 120$ & $ 73.9 \pm  6.0 $ \nl
PG\,1307 & \Hb\,\tablenotemark{b} & $108^{+ 46}_{-115}$ & $ 32^{+ 49}_{ -96}$ & 0.59  & 23 & $4190\pm210$ & $5260\pm 270$ & $ 19.2\pm  2.6$  \nl
PG\,1307 & \Hg & $201^{+ 72}_{-168}$ & $252^{+105}_{-426}$ & 0.64 & 23 & $4030\pm240$ & $3400\pm 200$ & $  7.4 \pm  1.3$ \nl
PG\,1351 & \Ha & $247^{+162}_{ -78}$ & $258^{+286}_{ -76}$ & 0.49 & 29 & $1170\pm160$ & $ 950\pm 130$ & $ 76.9 \pm  3.2 $ \nl
PG\,1411 & \Ha & $103^{+ 40}_{ -37}$ & $ 71^{+ 17}_{ -18}$ & 0.87 & 24 & $2172\pm 57$ & $2135\pm  56$ & $133.0 \pm  4.6 $ \nl
PG\,1411 & \Hb & $118^{+ 72}_{ -71}$ & $ 67^{+157}_{ -31}$ & 0.64 & 24 & $2456\pm 96$ & $2740\pm 110$ & $ 33.3\pm  1.8 $  \nl
PG\,1426 & \Ha & $ 90^{+ 46}_{ -52}$ & $ 66^{+ 56}_{ -59}$ & 0.87 & 20 & $5450\pm150$ & $4850\pm 130$ & $136.2\pm  9.6 $ \nl
PG\,1426 & \Hb & $115^{+ 49}_{ -68}$ & $ 61^{+ 81}_{ -66}$ & 0.83 & 20 & $6250\pm390$ & $5520\pm 340$ & $ 30.9 \pm  2.9$ \nl
PG\,1613 & \Ha & $ 43^{+ 40}_{ -22}$ & $ 35^{+185}_{ -26}$ & 0.43 & 44 & $6060\pm210$ & $1598\pm  56$ & $ 97.0 \pm  5.8 $\nl
PG\,1613 & \Hb & $ 44^{+ 20}_{ -23}$ & $ 29^{+ 36}_{ -31}$ & 0.43 & 48 & $7000\pm380$ & $2500\pm 140$ & $ 20.2\pm  1.3 $ \nl
PG\,1617 & \Ha & $111^{+ 31}_{ -37}$ & $ 83^{+ 45}_{ -32}$ & 0.80 & 32 & $4270\pm410$ & $3160\pm 300$ & $ 50.4 \pm  3.9 $ \nl
PG\,1617 & \Hb & $ 78^{+ 30}_{ -41}$ & $ 83^{+ 32}_{ -55}$ & 0.75 & 34 & $5120\pm850$ & $3880\pm 650$ & $ 14.0 \pm  1.7 $ \nl
PG\,1617 & \Hg & $ 43^{+117}_{ -58}$ & $ 50^{+148}_{ -49}$ & 0.67 & 32 & $3100\pm750$ & $3110\pm 750$ & $  4.2 \pm  1.1$ \nl
PG\,1700 & \Hb & $114^{+246}_{-235}$\,\tablenotemark{a} & $ 14^{+296}_{-177}$ & 0.54 & 37 & $2180\pm170$ & $1970\pm 150$ & $ 18.88\pm  0.99$ \nl
PG\,1704 & \Hb & $437^{+252}_{-391}$ & $527^{+263}_{-407}$ & 0.43 & 24 & $ 890\pm280$ & $ 400\pm 120$ & $  3.18\pm  0.41$ \nl
PG\,2130 & \Ha & $237^{+ 53}_{ -28}$ & $194^{+252}_{ -33}$ & 0.59 & 64 & $2022\pm 88$ & $2236\pm  97$ & $168\pm 11$  \nl
PG\,2130 & \Hb & $188^{+136}_{ -27}$ & $186^{+144}_{-103}$ & 0.66 & 64 & $2410\pm150$ & $3010\pm 180$ & $ 51.2 \pm  4.6$ \nl
PG\,2130 & \Hg & $196^{+130}_{ -46}$ & $218^{+104}_{-102}$ & 0.72 & 54 & $2850\pm190$ & $3050\pm 210$ & $ 22.7\pm  2.7 $ \nl
\enddata
%
% \tablenotetext{a}{The center of mass for the \Hb \ ICCF of PG0844 and
% PG1700 was computed from all points within 60\% of the peak value,
% $r_{max}$, due to noisiness of the data.} 
% \tablenotetext{b}{In the \Hb \ ICCF of PG~1307 we measured the right
% peak since it is the one consistent with the ZDCF results and the \Ha
% \ and \Hg \ of this object.}
% \tablenotetext{c}{For the \Hb \ and \Hg \ of PG~1617 and \Ha \ of
% PG~2130 the errors quoted were computed for the main peak that appeared
% in the CCPD.}
%
\tablenotetext{a}{Computed from all points within 60\% of the peak value,
$r_{max}$, due to the noisiness of the data.} 
\tablenotetext{b}{The peak of the ICCF measured is
the one consistent with the ZDCF results and with the \Ha\ and \Hg\ results.}
\end{deluxetable}
\normalsize

All emission-line light curves presented in Fig.~\ref{flcs} were cross
correlated with their corresponding continuum light curves.  The
results of the cross-correlation analysis are presented in
Table~\ref{tccfs}. The object name is given in Column (1) and the
specific Balmer line is listed in column (2). For the ICCF we list the
centroid time-lag, $\tau_{cent}$, (computed from all points within 80\%
of the peak value, $r_{max}$ \footnote{Koratkar \& Gaskell (1991a) show
the effects of calculating the centroid at various levels of the peak
value. The ICCFs presented here are smoother than those presented in
their study.  Therefore, a level chosen at 80\% of the peak should be
adequate.}) in column (3).  We define the ICCF peak as the point of
maximum correlation and list the peak position, $\tau_{peak}$, in
column (4), and the peak value, $r_{max}$, in column (5). In column (6)
we list the total number of points in the emission-line light curve
(the total number of points in the continuum light curve is given in
Table~\ref{tsample} columns [8] \& [9]). The uncertainties given for
$\tau_{cent}$ and $\tau_{peak}$ were computed with the FR/RSS method
using $\sim 10,000$ Monte Carlo realizations which were used to build
up a cross-correlation peak distribution (CCPD; Maoz \& Netzer 1989).
The range of uncertainties contains 68\% of the Monte Carlo
realizations in the CCPD and thus would correspond to 1$\sigma$
uncertainties for a normal distribution. In addition to the
cross-correlation results, we also list in Table~\ref{tccfs} velocity
measurements and the average of the observed flux for each line. Column (7)
lists the mean full-width at half maximum (FWHM) of the emission-line,
column (8) lists the line FWHM measured from the rms spectrum (see
\S~\ref{secmass} below), and column (9) lists the average of the observed line
flux and its rms (calculated from the light curves of
Table~\ref{linetab}).

\begin{figure*}
\vglue -1,9in
\centerline{\epsfxsize=21cm\epsfbox{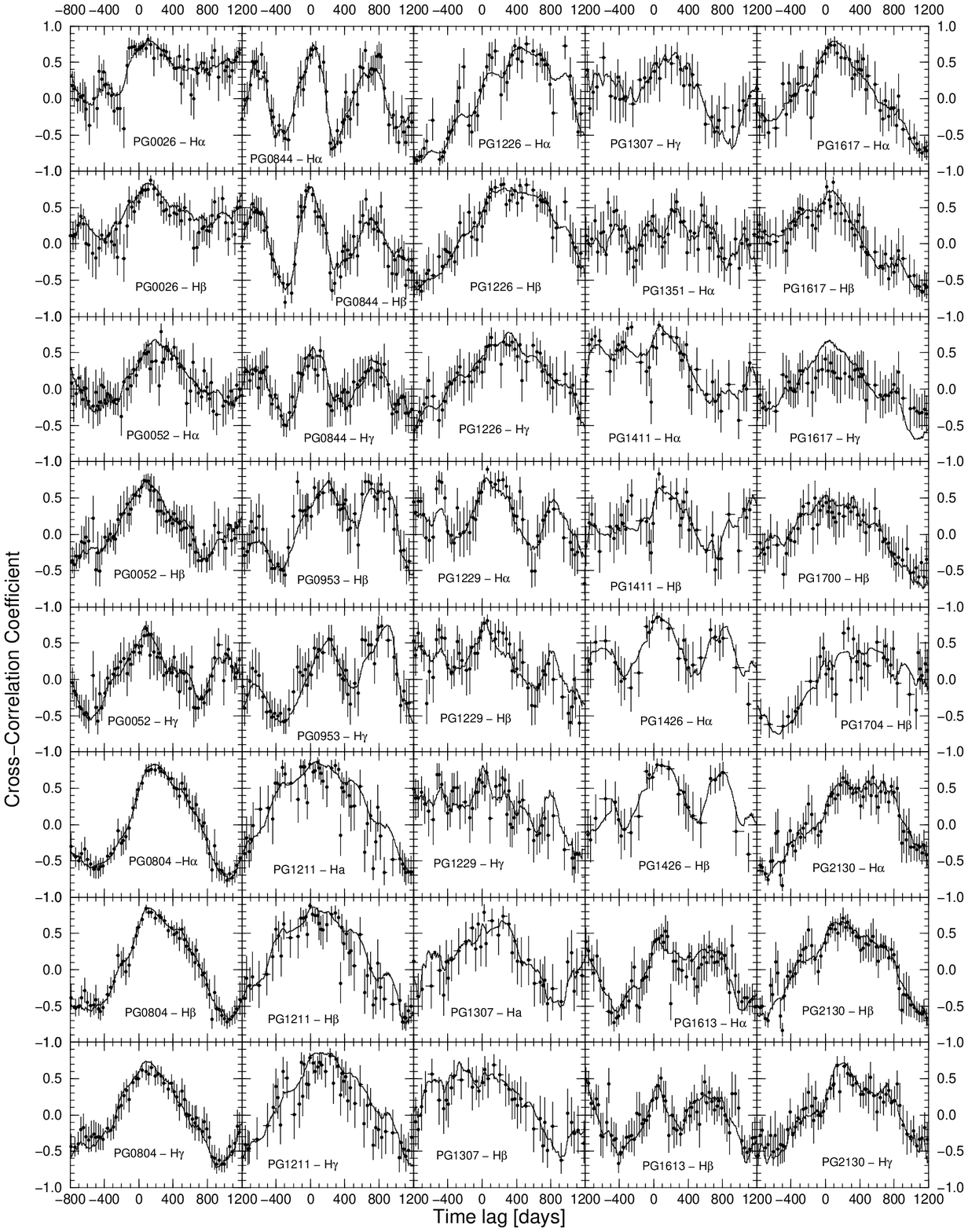}}
\vglue -0.8in
\caption{Cross correlation functions: ICCF {\it (solid line)} and ZDCF
{\it (circles with error bars)} for all emission lines with peak
correlation coefficients $>0.4$. See text and
Table~\protect{\ref{tccfs}} for details.}
\label{fccfs}
\end{figure*}

Fig.~\ref{fccfs} presents the cross-correlation functions (CCFs) for
the 40 out of 46 emission-line/continua pairs having peak correlation
coefficients above 0.4\,. All 40 CCFs indicate a positive time lag of
the Balmer lines with respect to the optical continuum. The time lags
are of order a few weeks to a few months, and the CCF peaks are highly
significant for most lines. In a few cases, $\tau_{cent}$ and
$\tau_{peak}$ are consistent with a zero time lag and we attribute this
to the fact that these emission-line light curves are noisy, and to the
conservative error estimate of the FR/RSS method. Choosing our best lag
determination (see below), i.e., avoiding noisy results obtained from
weak lines (generally \Hg ), we conclude that a time lag has been
detected in one or more of the Balmer lines for {\em all} 17 quasars.

When comparing the \Hb\ time-lags for PG\,0804 and PG\,0953 obtained in
Paper~II to the ones deduced here, we find an increase of about 65\%
for both objects, although the results are consistent at the 2$\sigma$
level. The luminosities of both objects also increased during the
second half of the monitoring period. Changes in the time lag
between different observing seasons is seen in lower-luminosity AGN
(e.g., NGC\,4151 and NGC\,5548) and is thought to be a real effect,
rather than a result of measurement error. For instance, the BLR may
become larger when the object becomes brighter.

\vspace{1cm}

\section{Size, Luminosity, and Mass Determination}
\label{SML}

\vspace{0.5cm}

Converting the observable Balmer emission-line time lags, line widths,
and continuum fluxes into interesting physical parameters, namely, BLR
sizes, and AGN central masses and luminosities, is not straightforward.
We therefore discuss each in some detail below. We analyze our data
together with data for other AGN. Wandel et al. (1999) have uniformly
analyzed the reverberation mapping data of 17 Seyfert~1s, and deduced
time lags using the same techniques described in the previous section.
Combining their results with the present ones, we obtain, for the first
time, reliable Size--Mass--Luminosity relations for 34 AGN spanning over
4 orders of magnitudes in continuum luminosity. Our analysis is similar
to that of Wandel et al. (1999), and the few significant differences
are noted below.

We apply below linear regression analysis to the data.  Unless
otherwise noted, the method is the one described by Press et al.
(1992), which is based on an iterative process to minimize $\chi^2$,
taking into account the uncertainties in both coordinates. In some of
our data, the uncertainties are asymmetric (e.g., as a result of the
FR/RSS method and the use of the CCPD), and since the fitting method
does not account for this, we use the mean of the positive and negative
uncertainty estimates in a given coordinate for each data point.  We
find that similar results are obtained when using the Pearson linear
correlation and the Spearman rank-order correlation. We present below
only Pearson coefficients.

\newpage

\subsection{The Size of the BLR}
\label{rblr}

Caution must be exercised when using time lags to estimate BLR sizes 
because there are various assumptions and complications involved:
\begin{enumerate}
\item The CCF is sensitive to the characteristics of the continuum
variation, since it is a convolution of the transfer function and the
continuum autocorrelation function (Penston 1991; Peterson 1993).
Hence, there is a dependence of the peak and centroid of the CCF on the
nature of the continuum variability. An illustration of this dependence
is given by Netzer \& Maoz (1990).
\item A real change in BLR size and geometry, over the 7 year period,
is not out of the question. Using the time-lags measured here, and the
observed FWHMs, we deduce a dynamical (crossing) time of the order of
30 years for the more luminous objects and about a year for the lowest
luminosity Seyfert galaxies.
\item The measured time lag can be affected by the nonlinear response
of some  emission-line intensities to  continuum  variations.
Nonlinear responses are suggested by various studies (O'Brien, Goad,
\& Gondhalekar, 1995; Shields, Ferland, \& Peterson, 1995) and
discussed by Kaspi \& Netzer (1999).
\item The ionizing continuum variations may behave differently from the
observed optical continuum. Due to anisotropic continuum emission, the
emission line gas may be illuminated by an ionizing continuum that is
poorly represented by the observed optical continuum (e.g. Netzer
1987). Furthermore, variable beaming or obscuration may be different
along the different lines of sight from the continuum source to the
observer and to the gas. Most studies show the continuum varies in
phase with little or no time delay (albeit, with smaller amplitude at
longer wavelengths) over a wide wavelength range.  There are, however,
some exceptions (e.g., Peterson et al. 1998b) and it is clear that a
full understanding of the continuum emission processes is yet to be
obtained.  An extreme example is shown by Nandra et al. (1998) who
found that the X-ray and UV continuum are not strongly correlated at
zero time lag in NGC 7469, with some X-ray emission peaks lagging
behind the UV peaks by $\sim$4 days. Recently, Maoz, Edelson, \& Nandra
(2000) have demonstrated even weaker correlation at zero lag between
X-ray and optical variations in NGC~3516.
\item For a given object, different lines from different ions have
different time lags. Thus, considering the entire sample, the same
emission lines should be used in all objects.  Balmer lines are useful
since their lag has been measured in most AGN where reverberation
mapping has been attempted. However, as argued by Kaspi \& Netzer
(1999), the theoretical modeling of these lines is very uncertain and
they are useful mostly because they provide a uniform base for
comparing low luminosity Seyferts with high luminosity quasars.
\item  There may be flux contributions by other emission lines blended
with the Balmer lines.  These emission lines may have a different time
lag and affect the measured emission line lag.  An example of this
potential situation is the variable Fe\,{\sc ii} emission lines near
\Hb .
\end{enumerate}

A possible manifestation of some of these difficulties is the fact that
for some objects, different monitoring campaigns  result in different
time lags. The effect was first noted in the \Hb\ time lags found by
Peterson et al. (1991; $\sim$7 days) and Netzer et al. (1990; $\sim$20
days) for the Seyfert galaxy NGC 5548. Other examples are the longer
time lag deduced for PG\,0804 and PG\,0953 in this paper, compared with
Paper~II (see above), and the different time lags in each observing
season for NGC~5548 (Peterson et al. 1999).

Bearing in mind the above points, we convert the time lags (deduced
from the ICCF centroid, $\tau_{cent}$ from Table~\ref{tccfs}) of the
Balmer emission lines directly into BLR sizes (hereafter \Rblr) after
applying a cosmological $(1+z)^{-1}$ factor. We find that time lags
derived from multiple Balmer lines are consistent with each other for
individual quasars.  The derived linear relations are:
\begin{equation}
R_{BLR} (H\alpha)=(1.19 \pm 0.23) R_{BLR}(H\beta)+ (13 \pm 19) \ \ ,
\end{equation}
and
\begin{equation}
R_{BLR}(H \gamma)=(0.96 \pm 0.30) R_{BLR}(H\beta) -(3 \pm 33) \ \ ,
\end{equation}
i.e., both are consistent with a slope of 1.0 and a zero intercept. In
view of this, we adopt the mean time lag measured for \Ha\ and \Hb\ as
the best estimate of \Rblr\,\footnote{When using only \Hb\ for the
analysis, we derive the same results but with somewhat larger
scatter.}. The relatively low equivalent width of \Hg\ makes it more
difficult to measure and we do not include it in the estimate of BLR
size. In several objects only one of \Ha\ or \Hb\ is available and we
thus use a single line. For the Seyfert~1s we use the \Hb\ time lags
from Wandel et al. (1999) and correct them by the $(1+z)^{-1}$ factor.
We have used the same method for calculating the time lags as that
used by Wandel et~al. (1999). Values for \Rblr\ for all
objects are listed in Table~\ref{trlm}, column~(2).

%\scriptsize
\begin{deluxetable}{lcccc}
\tablecolumns{5}
\tablewidth{0pt}
\tablecaption{Radii, Luminosities, \&  Masses
\label{trlm}}
\tablehead{
\colhead{Object} 		&
\colhead{$R_{BLR}$} 	&
\colhead{$\lambda L_{\lambda}$(5100\AA )} &
\colhead{$M$(mean)} 		&
\colhead{$M$(rms)} 		\nl
\colhead{} 		&
\colhead{(lt-days)} 	&
\colhead{$10^{44}$\,ergs\,s$^{-1}$} &
\colhead{$10^7M_{\odot}$}	&
\colhead{$10^7M_{\odot}$}	\nl
\colhead{(1)} & 
\colhead{(2)} & 
\colhead{(3)} &
\colhead{(4)} &
\colhead{(5)} 
} 
\startdata
3C\,120      &             $ 42^{+ 27}_{- 20}$ & $ 0.73\pm  0.13$ & $ 2.3^{+ 1.5}_{- 1.1}$ & $ 3.0^{+ 1.9}_{- 1.4}$ \nl
3C\,390.3    &       $ 22.9^{+  6.3}_{-  8.0}$ & $ 0.64\pm  0.11$ & $34^{+11}_{-13}$ & $37^{+12}_{-14}$ \nl
Akn\,120     &       $ 37.4^{+  5.1}_{-  6.3}$ & $ 1.39\pm  0.26$ & $18.4^{+ 3.9}_{- 4.3}$ & $18.7^{+ 4.0}_{- 4.4}$ \nl
F\,9         &       $ 16.3^{+  3.3}_{-  7.6}$ & $ 1.37\pm  0.15$ & $ 8.0^{+ 2.4}_{- 4.1}$ & $ 8.3^{+ 2.5}_{- 4.3}$ \nl
IC\,4329A    &     $  1.4^{+  3.3}_{-  2.9}$ & $ 0.164\pm  0.021$ & $ 0.5^{+ 1.3}_{- 1.1}$ & $ 0.7^{+ 1.8}_{- 1.6}$ \nl
Mrk\,79      &     $ 17.7^{+  4.8}_{-  8.4}$ & $ 0.423\pm  0.056$ & $ 5.2^{+ 2.0}_{- 2.8}$ & $10.2^{+ 3.9}_{- 5.6}$ \nl
Mrk\,110     &       $ 18.8^{+  6.3}_{-  6.6}$ & $ 0.38\pm  0.13$ & $ 0.56^{+ 0.20}_{- 0.21}$ & $ 0.77^{+ 0.28}_{- 0.29}$ \nl
Mrk\,335     &     $ 16.4^{+  5.1}_{-  3.2}$ & $ 0.622\pm  0.057$ & $ 0.63^{+ 0.23}_{- 0.17}$ & $ 0.38^{+ 0.14}_{- 0.10}$ \nl
Mrk\,509     &       $ 76.7^{+  6.3}_{-  6.0}$ & $ 1.47\pm  0.25$ & $ 5.78^{+ 0.68}_{- 0.66}$ & $ 9.2^{+ 1.1}_{- 1.1}$ \nl
Mrk\,590     &     $ 20.0^{+  4.4}_{-  2.9}$ & $ 0.510\pm  0.096$ & $ 1.78^{+ 0.44}_{- 0.33}$ & $ 1.38^{+ 0.34}_{- 0.25}$ \nl
Mrk\,817     &     $ 15.0^{+  4.2}_{-  3.4}$ & $ 0.526\pm  0.077$ & $ 4.4^{+ 1.3}_{- 1.1}$ & $ 3.54^{+ 1.03}_{- 0.86}$ \nl
NGC\,3227    &   $ 10.9^{+  5.6}_{- 10.9}$ & $ 0.0202\pm  0.0011$ & $ 3.9^{+ 2.1}_{- 3.9}$ & $ 4.9^{+ 2.6}_{- 4.9}$ \nl
NGC\,3783    &     $  4.5^{+  3.6}_{-  3.1}$ & $ 0.177\pm  0.015$ & $ 0.94^{+ 0.92}_{- 0.84}$ & $ 1.10^{+ 1.07}_{- 0.98}$ \nl
NGC\,4051    & $  6.5^{+  6.6}_{-  4.1}$ & $ 0.00525\pm  0.00030$ & $ 0.13^{+ 0.13}_{- 0.08}$ & $ 0.14^{+ 0.15}_{- 0.09}$ \nl
NGC\,4151    &   $  3.0^{+  1.8}_{-  1.4}$ & $ 0.0720\pm  0.0042$ & $ 1.53^{+ 1.06}_{- 0.89}$ & $ 1.20^{+ 0.83}_{- 0.70}$ \nl
NGC\,5548    &     $ 21.2^{+  2.4}_{-  0.7}$ & $ 0.270\pm  0.053$ & $12.3^{+ 2.3}_{- 1.8}$ & $ 9.4^{+ 1.7}_{- 1.4}$ \nl
NGC\,7469    &     $  4.9^{+  0.6}_{-  1.1}$ & $ 0.553\pm  0.016$ & $ 0.65^{+ 0.64}_{- 0.65}$ & $ 0.75^{+ 0.74}_{- 0.75}$ \nl
PG\,0026 &               $113^{+ 18}_{- 21}$ & $ 7.0\pm  1.0$ & $ 5.4^{+ 1.0}_{- 1.1}$ & $ 2.66^{+ 0.49}_{- 0.55}$ \nl
PG\,0052 &               $134^{+ 31}_{- 23}$ & $ 6.5\pm  1.1$ & $22.0^{+ 6.3}_{- 5.3}$ & $30.2^{+ 8.8}_{- 7.4}$ \nl
PG\,0804 &               $156^{+ 15}_{- 13}$ & $ 6.6\pm  1.2$ & $18.9^{+ 1.9}_{- 1.7}$ & $16.3^{+ 1.6}_{- 1.5}$ \nl
PG\,0844 &       $ 24.2^{+ 10.0}_{-  9.1}$ & $ 1.72\pm  0.17$ & $ 2.16^{+ 0.90}_{- 0.83}$ & $ 2.7^{+ 1.1}_{- 1.0}$ \nl
PG\,0953 &               $151^{+ 22}_{- 27}$ & $11.9\pm  1.6$ & $18.4^{+ 2.8}_{- 3.4}$ & $16.4^{+ 2.5}_{- 3.0}$ \nl
PG\,1211 &             $101^{+ 23}_{- 29}$ & $ 4.93\pm  0.80$ & $ 4.05^{+ 0.96}_{- 1.21}$ & $ 2.36^{+ 0.56}_{- 0.70}$ \nl
PG\,1226 &               $387^{+ 58}_{- 50}$ & $64.4\pm  7.7$ & $55.0^{+ 8.9}_{- 7.9}$ & $23.5^{+ 3.7}_{- 3.3}$ \nl
PG\,1229 &             $ 50^{+ 24}_{- 23}$ & $ 0.94\pm  0.10$ & $ 7.5^{+ 3.6}_{- 3.5}$ & $ 8.6^{+ 4.1}_{- 4.0}$ \nl
PG\,1307 &             $124^{+ 45}_{- 80}$ & $ 5.27\pm  0.52$ & $28^{+11}_{-18}$ & $33^{+12}_{-22}$ \nl
PG\,1351 &             $227^{+149}_{- 72}$ & $ 4.38\pm  0.43$ & $ 4.6^{+ 3.2}_{- 1.9}$ & $ 3.0^{+ 2.1}_{- 1.3}$ \nl
PG\,1411 &             $102^{+ 38}_{- 37}$ & $ 3.25\pm  0.28$ & $ 8.0^{+ 3.0}_{- 2.9}$ & $ 8.8^{+ 3.3}_{- 3.2}$ \nl
PG\,1426 &             $ 95^{+ 31}_{- 39}$ & $ 4.09\pm  0.63$ & $47^{+16}_{-20}$ & $37^{+13}_{-16}$ \nl
PG\,1613 &             $ 39^{+ 20}_{- 14}$ & $ 6.96\pm  0.87$ & $24.1^{+12.5}_{- 8.9}$ & $ 2.37^{+ 1.23}_{- 0.88}$ \nl
PG\,1617 &             $ 85^{+ 19}_{- 25}$ & $ 2.37\pm  0.41$ & $27.3^{+ 8.3}_{- 9.7}$ & $15.4^{+ 4.7}_{- 5.5}$ \nl
PG\,1700 &               $ 88^{+190}_{-182}$ & $27.1\pm  1.9$ & $ 6^{+13}_{-13}$ & $ 5.0^{+11}_{-10}$ \nl
PG\,1704 &               $319^{+184}_{-285}$ & $35.6\pm  5.2$ & $ 3.7^{+ 3.1}_{- 4.0}$ & $ 0.75^{+ 0.63}_{- 0.81}$ \nl
PG\,2130 &             $200^{+ 67}_{- 18}$ & $ 2.16\pm  0.20$ & $14.4^{+ 5.1}_{- 1.7}$ & $20.2^{+ 7.1}_{- 2.4}$ \nl
\enddata
\end{deluxetable}
\normalsize

\begin{figure*}[t]
\centerline{\epsfxsize=8.5cm\epsfbox{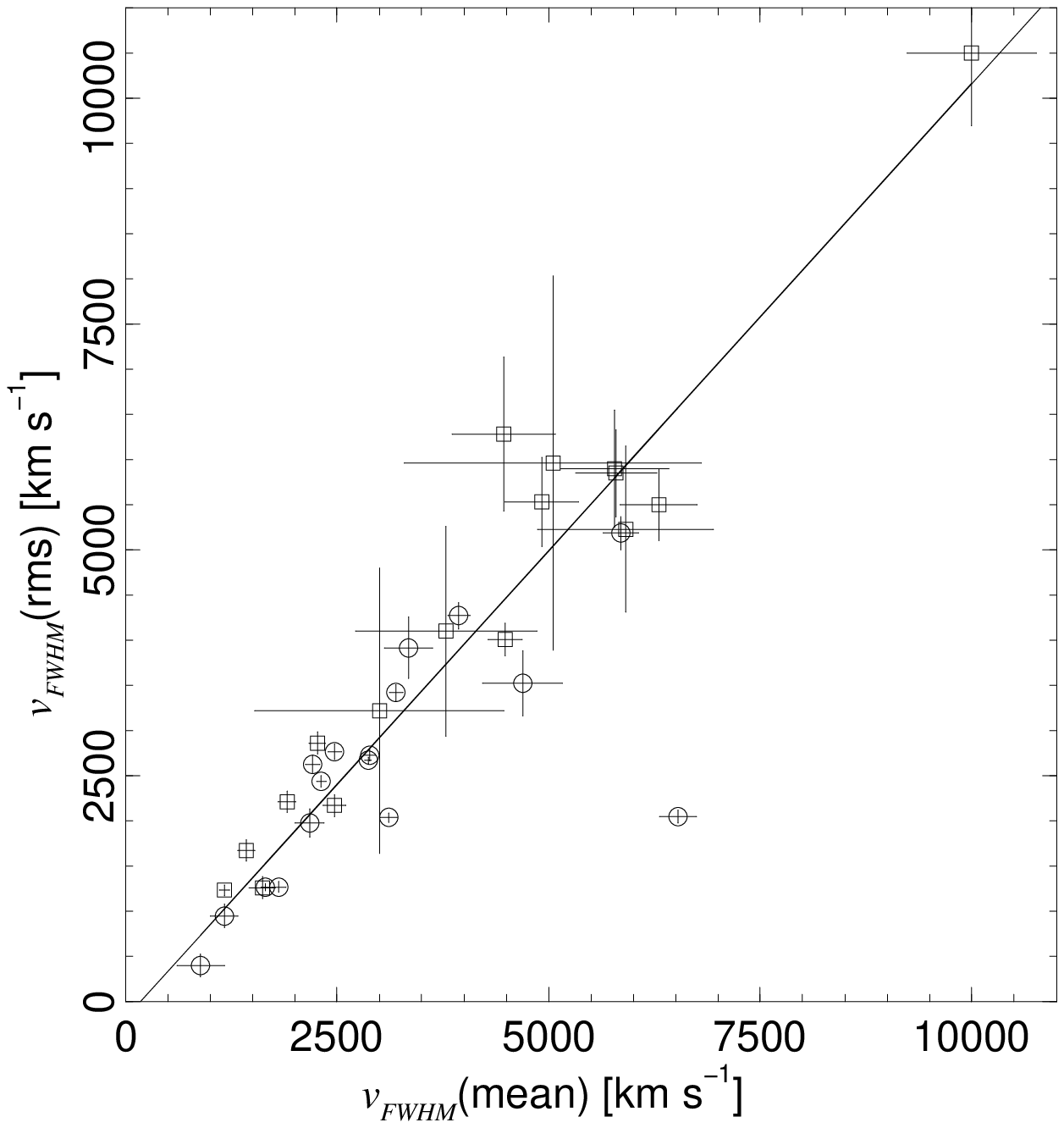}}
\caption{$v_{FWHM}(rms)$ vs. $v_{FWHM}(mean)$. 
In this and subsequent figures, PG quasars are denoted by circles and
Seyfert 1s are denoted by squares. The solid line
is the best fit to the data. The outlier is PG\,1613, which apparently
varies only in the core of its very broad lines. The object with the
broadest lines is 3C\,390.3\,.}
\label{ffvsf}
\end{figure*}

\subsection{Luminosity Determination}
\label{seclum}

The exact shape of the spectral energy distribution (SED) of most
objects in our sample is unknown. Furthermore, much of this luminosity
is emitted in the unobservable far UV and there are still unsolved
fundamental issues concerning the shape of the continuum (e.g., Zheng
et al. 1997, Laor et al. 1997, and references therein). Even in those
cases where multiwavelength data exist, the data are generally obtained
at different epochs, and variability may render a simple integration
over a wide wavelength range inappropriate. (See also a discussion by
Dumont, Collin-Souffrin, \& Nazarova; 1998, that demonstrates how
poorly the SED is determined for even one of the best-studied AGN, NGC
5548.) Intrinsic reddening and possible orientation effects complicate
the situation and it is thus unclear what fraction of the total
luminosity we are sampling.

An additional complication is the contribution of the host galaxy to
the luminosity of the nucleus. While in high-luminosity AGN this
contribution is negligible, this is not the case in low-luminosity
objects.  Determining the host galaxy flux within a given spectrograph
aperture is a complicated task, attempted only in a few cases (e.g.,
Peterson et al. 1995). The implementation of such corrections is beyond
the scope of this paper.

Following Wandel et al. (1999), we  use $\lambda L_{\lambda}$ at 5100
\AA \ (rest frame) as our luminosity measure. This luminosity is
calculated assuming zero cosmological constant, deceleration parameter
$q_0=0.5$, and Hubble constant $H_0=75$\,km\,s$^{-1}$Mpc$^{-1}$ (note
that these values are different from the ones used by Wandel et al.
1999). This monochromatic luminosity can be translated into an
integrated 0.1--1 $\mu$m luminosity, assuming a power-law continuum of
the form $f_{\nu}\propto\nu^{-0.5}$, as in Papers I \& II, by applying
a correction factor of 3.1. For the 5100 \AA\ point, we use the
continuum measurement on the red side of \Hb\ (see wavelength ranges in
column~(7) of Table~\ref{tabwave}). The uncertainty in this value is
taken to be only due to the variation range of each object, and is
represented by the rms of the light curve. The observed flux has been
corrected for Galactic extinction using the $A_B$ values from
NED\footnote{The NASA/IPAC Extragalactic Database (NED) is operated by
the Jet Propulsion Laboratory, California Institute of Technology,
under contract with the National Aeronautics and Space Administration.}
(see Table~\ref{tsample} column~[7]) and a standard extinction curve
(Savage \& Mathis 1979). Monochromatic luminosities, $\lambda
L_{\lambda}$(5100 \AA ), are listed in Table~\ref{trlm}, column~(3).

We have also tried an alternative process to determine the optical
luminosity. Using line-free bands in the optical spectra, we have
fitted a continuum to the mean and rms spectra of all objects. We
integrated those continua over the rest frame wavelength of
4500--6000~\AA \ to obtain the mean luminosity and its range. We have
found the results to be very similar to those obtained by using the
monochromatic luminosity.  We have also examined the consequences of
using the \Hb\ luminosity (rather than the optical luminosity) to
represent the ionizing luminosity of each AGN. Here too, we have found
the results presented below to be mostly unchanged by this choice.
Below we use only $\lambda L_{\lambda}(5100$ \AA ) as our luminosity
measure.

\subsection{Mass Determination}
\label{secmass}

To estimate the central masses of the quasars we need to assume
gravitationally dominated motions of the BLR clouds ($M\approx
G^{-1}v^2r$). This assumption is presented in Gaskell (1988) for the
Seyfert 1 galaxy NGC\,4151 and is also discussed and justified by
Wandel et al. (1999) and Peterson \& Wandel (1999) for the Seyfert 1
galaxy NGC~5548 (though see Gaskell 1996 for a more general approach).
There are, however, several complications regarding the velocity field.
Although it is generally believed that the line widths represent a
Keplerian velocity, it is not clear how to convert the observed profile
into a velocity measure, since the observed profiles are different from
the ones expected from simple Keplerian orbits of arbitrary
inclinations. In the absence of such information, it is common to use
the rest-frame FWHM of the emission line. Because the broad emission
lines of AGN are composed of a narrow component superposed on a broader
components, a unique FWHM determination is not straightforward.  In
particular, deblending the components is complicated since the line
profiles cannot be represented by simple analytical functions.  Also,
the line profiles and widths are variable as is evident, for example,
in Gilbert et~al. (1999) and from the spectra obtained in this study.
These variations can be due to a real change in the BLR velocity field,
or other reasons, such as blending with weak lines, or observational
effects (e.g., inaccurate wavelength calibration, or changes in the
seeing).

\begin{figure*}[t]
\centerline{\epsfxsize=8.5cm\epsfbox{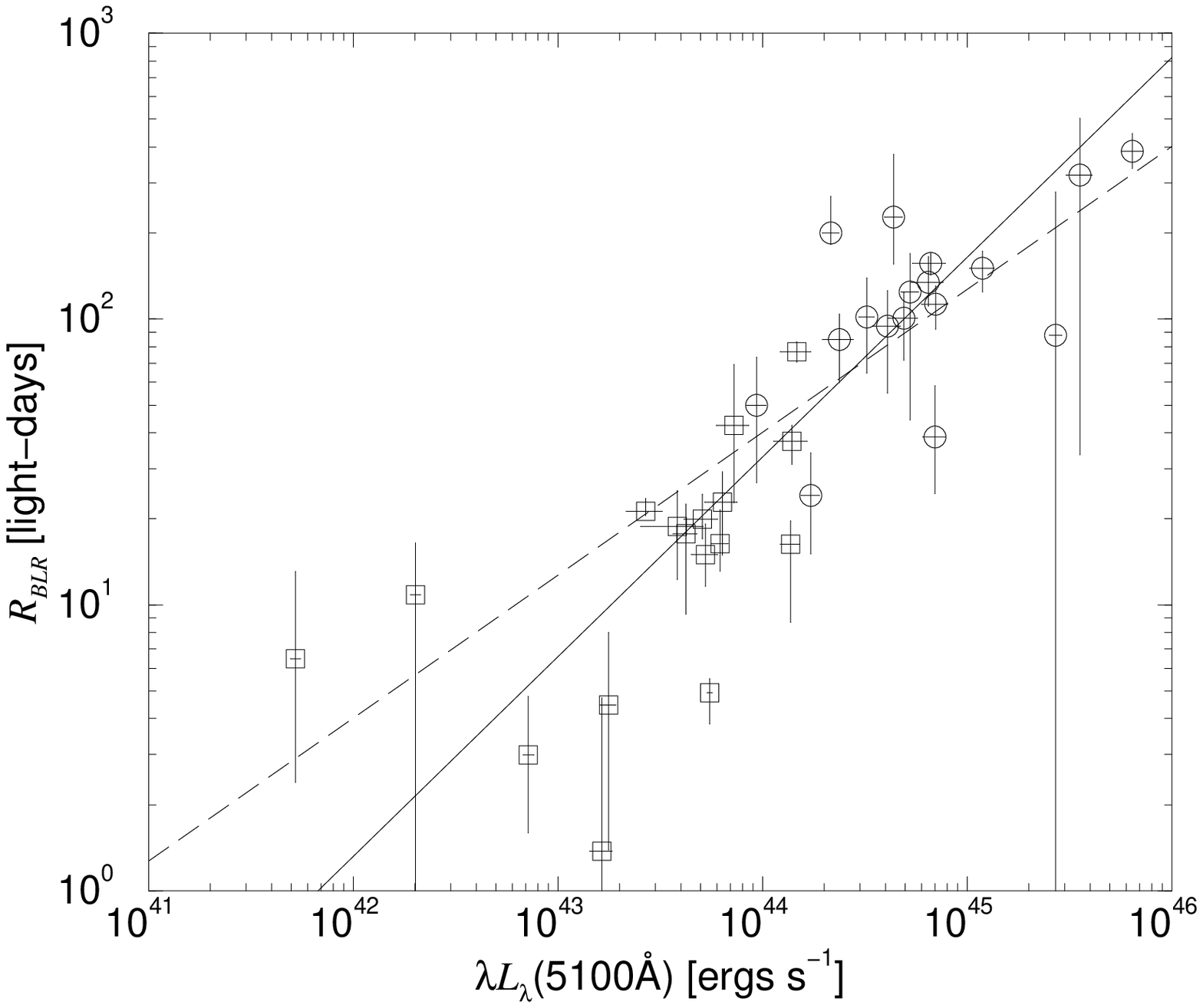}}
\caption{BLR size---luminosity relation. The solid line is the best
fit to the data. The dashed line is a fit with a slope of 0.5\,.}
\label{frvsl}
\end{figure*}

We use two different methods to estimate a typical velocity. In the
first method, for each quasar we measure the FWHM of each line in all
the spectra and calculate the mean FWHM, $v_{FWHM}$(mean), and its
uncertainty (rms). This empirical way of estimating the scatter
accounts for all the uncertainties involved, such as spectral
resolution, pixelation of the line profile, and uncertainties in
continuum determination.  The values are listed in Table~\ref{tccfs},
column (7).

The second method, proposed by Peterson et al. (1998a), uses the rms
spectrum to compute the FWHM of the lines. In principle, constant
features in the mean spectrum (such as narrow forbidden emission lines,
narrow components of the permitted emission lines, galactic
absorptions, and constant continuum and broad-line features) are
excluded in this method. The FWHM measures only the part of the line
that varied and thus corresponds to the \Rblr\ measured from the
reverberation mapping\footnote{The rms spectra shown by Peterson et al.
(1998a) often show narrow forbidden line residuals that are probably
due to variable seeing and small shifts in wavelength calibration. We
see the same effect in our rms spectra as well. It is likely that the
same effect operates on the narrow component of the Balmer lines, and
therefore, the rms spectra are also not completely free of constant
narrow components.  The degree of this contamination may vary from
object to object.}.  For the rms spectrum of each object we have listed
the emission lines' FWHM, $v_{FWHM}$(rms), in  column (8) of
Table~\ref{tccfs}. There is no simple way to determine the uncertainty
of $v_{FWHM}$(rms), and instead we have used the relative uncertainty
in $v_{FWHM}$(mean).

Like the time lags derived using \Ha\ and \Hb , the FWHM of all lines
for a given object are consistent with each other.  Thus, for the PG
quasars we use the average FWHM of \Ha\ and \Hb. For the Seyfert 1
galaxies we have used the mean and rms FWHM quoted in Wandel et al.
(1999). The FWHM from the rms spectrum versus the mean FWHM from all
spectra is plotted in Fig.~\ref{ffvsf}.  We find the correlation to be
highly significant with correlation coefficient of 0.888\,.  A linear
fit gives the relation
\begin{equation}
v_{FWHM}(rms)=(1.035\pm0.034)v_{FWHM}(mean)-(185\pm93) \ \ ,  
\end{equation}
which is plotted as the solid line in Fig.~\ref{ffvsf}. The two
velocity estimates are similar and Peterson et al.'s (1998a) preference
for the FWHM determined from the rms spectra is not empirically
justified. However, in the subsequent analysis we will use both estimates
of the broad-line FWHM for each object.

To determine $v$ we correct $v_{FWHM}$ by a factor of
$\sqrt{3}/2$, to account for velocities in three dimensions and for
using half of the FWHM. The virial ``reverberation'' mass is then:
\begin{equation}
M=1.464\times 10^5\left(\frac{R_{BLR}}{\rm lt\,days}\right)
\left(\frac{v_{FWHM}}{\rm 10^3\,km\,s^{-1}}\right)^{2}M_{\odot} \ \ .
\end{equation} 
The results of this calculation are listed in columns (4) and (5) of
Table~\ref{trlm} for $v_{FWHM}$(mean) and $v_{FWHM}$(rms), respectively.

\vspace{0.4cm}

\section{Discussion}

\subsection {Size--Luminosity Relation}
\label{rvsl}

The \Rblr--luminosity relation is presented in Fig.~\ref{frvsl}. The
correlation coefficient is 0.827, and its significance level is
1.7$\times10^{-9}$. A linear fit to the points gives
%\vspace{-0.1cm}
\begin{equation}
%
%  \log(R_{BLR})=(0.699\pm 0.033)\log(\lambda L_{\lambda}(5100\,\AA )) - 
% (29.2\pm 1.5)
%
% R_{BLR} = 10^{-29.2\pm 1.5} (\lambda L_{\lambda}(5100\AA ))^{(0.699\pm 0.033)}
%
% R_{BLR} = (36^{+15}_{-11}\,lt\,days)\left(\frac{\lambda L_{\lambda}(5100\AA 
% )}{10^{44}\,ergs\,s^{-1}}\right)^{(0.699\pm 0.033)}
%
% R_{BLR} = (36\,lt\,days)\left(\frac{\lambda L_{\lambda}(5100\AA 
% )}{10^{44}\,ergs\,s^{-1}}\right)^{(0.699\pm 0.033)}
%
% R_{BLR} = 36\left(\frac{\lambda L_{\lambda}(5100\AA 
% )}{10^{44}\,ergs\,s^{-1}}\right)^{0.699\pm 0.033}\,lt\,days
% 
R_{BLR} = \left(32.9^{+2.0}_{-1.9}\right)\left(\frac{\lambda
L_{\lambda}(5100\,{\mbox{\AA}} )}{\rm 10^{44}\,ergs\,s^{-1}}
\right)^{0.700\pm 0.033}\,{\rm lt\,days} 
\end{equation}
%\vspace{-0.1cm}
(solid line plotted in Fig.~\ref{frvsl}). Considering the Seyfert
nuclei ($\log(\lambda L_{\lambda}$(5100 \AA )$\ltorder 44.2$), or the
PG quasars alone, we find only marginally significant correlations,
probably because of the narrow luminosity ranges.  A significant
correlation emerges only when using the whole luminosity range.

The present result is remarkable for two reasons. First, earlier
studies of this kind found a smaller power-law index in the BLR
size--AGN luminosity relation (closer to 0.5, e.g., Koratkar \& Gaskell
1991b; Wandel et~al. 1999).  A line with this slope was fit to the data
and is shown as a dashed line in Fig.~\ref{frvsl}\,. Our result using
the combined sample is not consistent with previous results.  Second,
under the assumptions that the shape of the ionizing continuum in AGN
is independent of $L$, and that all AGN are characterized by the same
ionization parameter and BLR density (as suggested by the similar line
ratios in low- and high-luminosity sources), one expects
$R_{BLR}\propto L^{0.5}$\,. The theoretical prediction is based on the
assumption that the gas distribution, and hence the mean BLR size,
scales with the strength of the radiation field.  Our present result
suggests that those assumptions should be re-examined.  This is also
implied from recent models (e.g., Kaspi \& Netzer 1999 and references
therein) which show a wide distribution of properties (such as BLR
density, column density, and ionization parameter) across the BLR of a
single active nucleus.  Therefore, a range of properties may also exist
among different AGN, and the above assumptions of uniform ionization
parameter and BLR density for all AGN is likely incorrect.  If an
effective ionization parameter can be defined, our result suggests that
it may be a decreasing function of luminosity.

\begin{figure*}[t]
\centerline{\epsfxsize=8.5cm\epsfbox{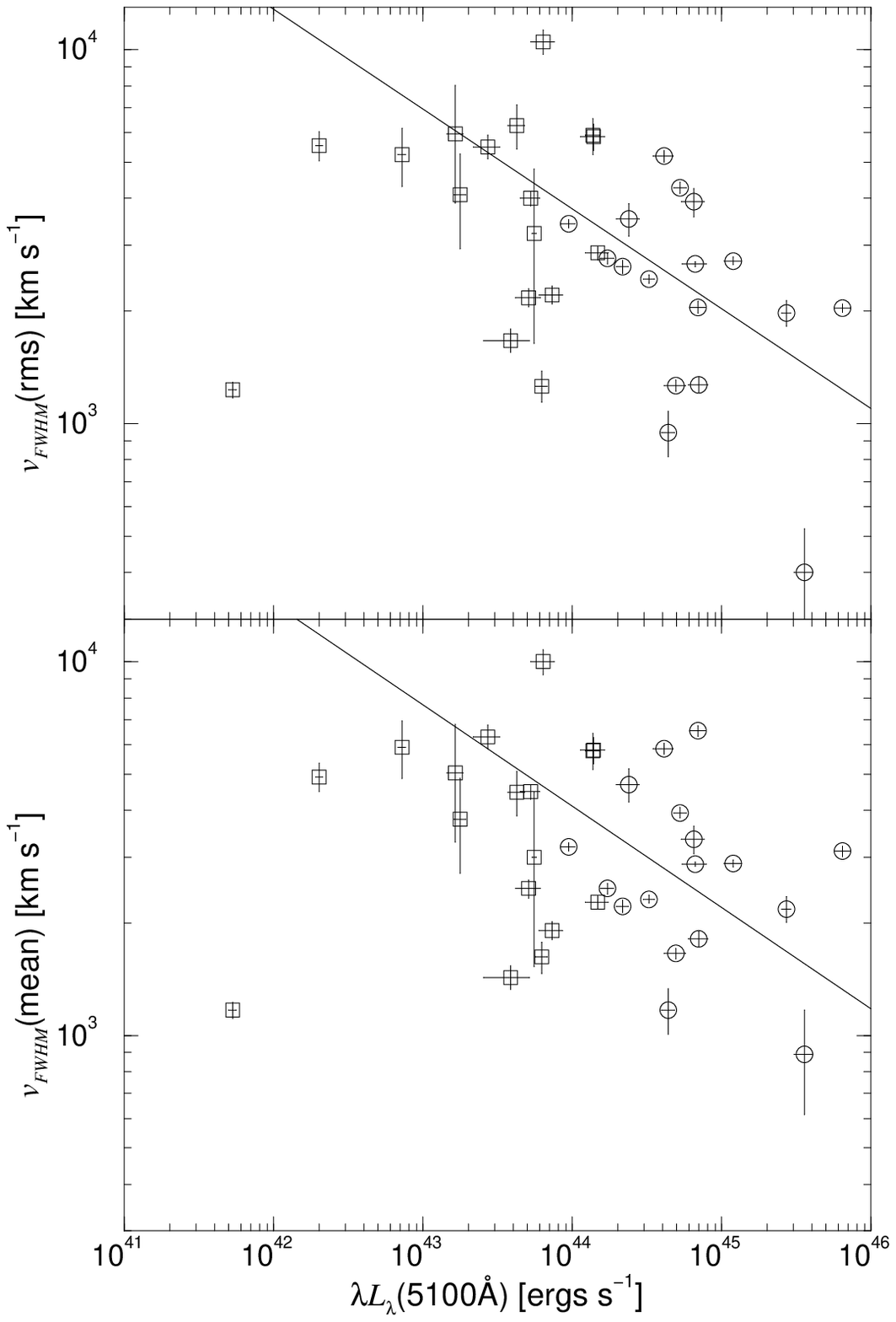}}
\caption{FWHM---luminosity relations. Top: FWHM derived from the rms
spectrum of each object. Bottom: FWHM derived from averaging the FWHMs
in all spectra for a given object. Solid lines are the best fit to the
data after excluding the outlier at the lowest luminosity (NGC~4051).}
\label{ffvsllog}
\end{figure*}

\subsection {FWHM--Luminosity Relation}
\label{vvsl}

Our sample allows us to to re-address the issue of the
velocity--luminosity relation in AGN.  Shuder (1984) noted that the
FWHM of the Balmer lines increases with luminosity for a sample of 25
AGN.  Wandel \& Yahil (1985) found a correlation coefficient of 0.5
between the \Hb\ full width at zero intensity (FW0I) and the
4000\,\AA\ continuum luminosity for a literature compilation of 94
AGN.  Joly et al. (1985; and references therein) also reported a weak
correlation between the FW0I of \Hb\ and the optical luminosity in a
similar collection of objects.  Boroson \& Green (1992) note the FW0I
of \Hb\ is quite sensitive to noise and depends strongly on the quality
of the Fe{\sc\,ii} subtraction. These authors found an anticorrelation
coefficient of $-$0.275 between the \Hb\ FWHM and absolute $V$
magnitude, significant at the 99\% confidence level, in a sample of 87
PG quasars. In a recent work Stirpe, Robinson, \& Axon (1999) measured
\Ha\ velocities for 126 AGN and found them to weakly correlate with the
luminosity.

\begin{figure*}[t]
\centerline{\epsfxsize=8.5cm\epsfbox{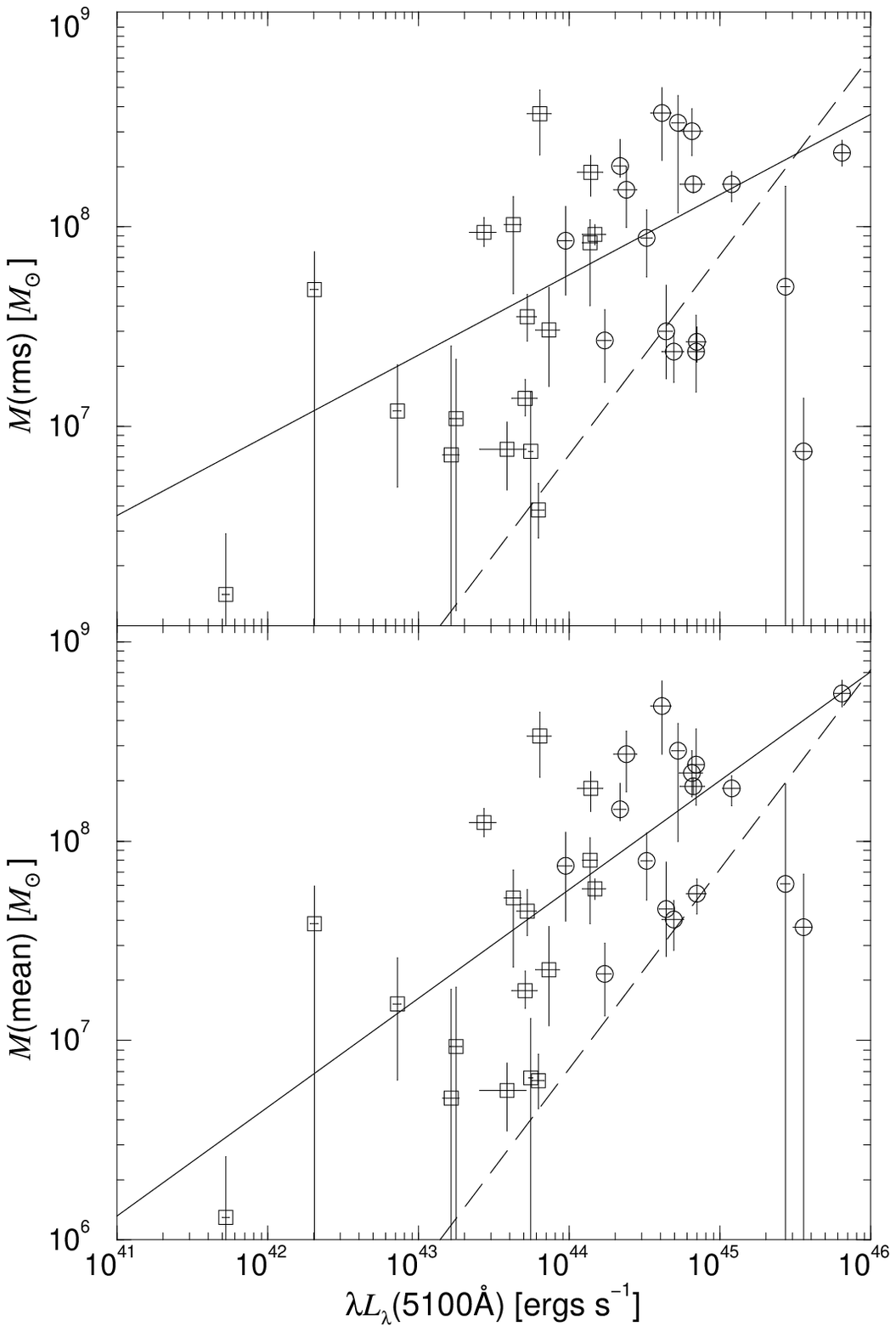}}
\caption{Mass---luminosity relations. Top: masses derived from
$v_{FWHM}$(rms). Bottom: masses derived from $v_{FWHM}$(mean).  Solid
lines are the best fit to the data. Dashed lines are the Eddington
limit based on a rough estimate for the bolometric luminosity (see text).}
\label{fmvsl}
\end{figure*}

In addressing the issue of the velocity--luminosity relation we note
that our sample is {\it not} complete, especially with the inclusion of
the heterogeneous Seyfert sample. The results must therefore be
treated with caution.  The FWHMs from the mean spectrum are plotted
against the luminosity in the bottom panel of Fig.~\ref{ffvsllog}.
There is no significant correlation between the variables. However, if
we omit the data point for the narrow line Seyfert~1, NGC~4051, which
clearly deviates from the other points and highly influences the fit,
we do obtain a marginally significant correlation.  The correlation
coefficient  is $-$0.386, and its significance level is 2.7$\times
10^{-2}$\,. A linear fit gives
\begin{equation}
%
% \log(v_{FWHM}(mean))=(-0.271\pm0.012)\log(\lambda L_{\lambda}(5100\,\AA 
% ))+(15.54\pm0.54)  \ \ .
%
% v_{FWHM}(mean) = 10^{(15.54\pm0.54)} (\lambda 
% L_{\lambda}(5100\,\AA))^{(-0.271\pm0.012)}
%
% v_{FWHM}(mean) = (4130\,km\,s^{-1})\left(\frac{\lambda L_{\lambda}(5100\AA 
% )}{10^{44}\,ergs\,s^{-1}}\right)^{(-0.271\pm0.012)}
%
v_{FWHM}(mean) = \left(4108^{+66}_{-65}\right)\left(\frac{\lambda L_{\lambda}(
5100{\mbox{\AA}})}{\rm 10^{44}\,ergs\,s^{-1}}\right)^{-0.271\pm0.012}\,{\rm km\,s^{-1}} \ \ .
\end{equation}
FWHMs from the rms spectrum versus luminosity are plotted in
the top panel of Fig.~\ref{ffvsllog}. The correlation coefficient is
$-$0.540 has a significance level of 1.2$\times 10^{-3}$ and the best linear
fit to the data is
\begin{equation}
%
% \log(v_{FWHM}(rms))=(-0.267\pm0.010)\log(\lambda L_{\lambda}(5100\,\AA 
% ))+(15.34\pm0.43) \ \ .
%
% v_{FWHM}(rms) = 10^{(15.34\pm0.43)} (\lambda 
% L_{\lambda}(5100\,\AA))^{(-0.267\pm0.010)}
%
% v_{FWHM}(rms) = (3910\,km\,s^{-1})\left(\frac{\lambda L_{\lambda}(5100\AA 
% )}{10^{44}\,ergs\,s^{-1}}\right)^{(-0.267\pm0.010)}
%
v_{FWHM}(rms) = \left(3752^{+57}_{-56}\right)\left(\frac{\lambda L_{\lambda}
(5100{\mbox{\AA}})}{\rm 10^{44}\,ergs\,s^{-1}}\right)^{-0.267\pm0.010}\,{\rm km\,s^{-1}}  \ \ .
\end{equation}
Thus, we find significant anticorrelations, in our incomplete sample,
between the FWHM of the Balmer lines and the luminosity of the objects
such that $v_{FWHM}\propto L^{-0.27\pm0.01}$.  Our result indicates a
stronger anticorrelation coefficient when $v_{FWHM}(rms)$ is used.
While the above studies used only one epoch for each quasar in their
measurements, our study uses the means of the luminosities and
velocities for each object over 7.5 years, and averages both \Ha\ and
\Hb . The PG quasars' $v_{FWHM}$ found in this study agrees with the
ones found by Boroson \& Green's (1992) to within $\sim$10\%, except
for two objects (PG\,1351 and PG\,1704) for which we measure
significantly narrower lines. If we exclude those two objects, the
anticorrelation becomes weaker and its significance decreases.  As the
correlation we find is opposite to those found by previous studies, the
issue of the velocity--luminosity relation needs further investigation,
which is beyond the scope of this paper.

\subsection {Mass--Luminosity Relation}

Our two velocity estimates produce two mass estimates for each object
(see Table~\ref{trlm} and \S~\ref{secmass}). Our mass estimates based
on the determination of the FWHM from the mean spectra are plotted
versus luminosity in the bottom panel of Fig.~\ref{fmvsl}. The
correlation coefficient between these two parameters is 0.646 and has a
significance level of 3.7$\times10^{-5}$. A linear fit gives
\begin{equation}
%
% \log(M(mean))=(0.546\pm0.036)\log(\lambda L_{\lambda}(5100\,\AA 
% ))+(17.0\pm1.6)
%
% M(mean) = 10^{(17.0\pm1.6)} (\lambda L_{\lambda}(5100\,\AA 
% ))^{(0.546\pm0.036)}
%
% M(mean) = (5.31\times10^7M_{\odot})\left(\frac{\lambda L_{\lambda}(5100\AA 
% )}{10^{44}\,ergs\,s^{-1}}\right)^{(0.546\pm0.036)}
%
% M(mean) = 5.31\times10^7\left(\frac{\lambda L_{\lambda}(5100\AA 
% )}{10^{44}\,ergs\,s^{-1}}\right)^{0.546\pm0.036}M_{\odot}
%
M(mean) = \left(5.71^{+0.46}_{-0.37}\right)\times10^7
\left(\frac{\lambda L_{\lambda}(5100\,{\mbox{\AA}} 
)}{\rm 10^{44}\,ergs\,s^{-1}}\right)^{0.545\pm0.036}M_{\odot}
\end{equation}
and is plotted as a solid line in the diagram.

The mass estimates based on the determination of the FWHM measured from the
rms spectra are plotted in the top panel of Fig.~\ref{fmvsl}. We find
correlation coefficient of 0.473 with a significance level of
4.7$\times10^{-3}$. A linear fit to this relation gives
\begin{equation}
%
% \log(M(rms))=(0.403\pm0.034)\log(\lambda L_{\lambda}(5100\,\AA ))+(23.4\pm1.5)
%
% M(rms) = 10^{(23.4\pm1.5)} (\lambda L_{\lambda}(5100\,\AA ))^{(0.403\pm0.034)}
%
% M(rms) = (6.81\times10^7M_{\odot})\left(\frac{\lambda L_{\lambda}(5100\AA 
% )}{10^{44}\,ergs\,s^{-1}}\right)^{(0.403\pm0.034)}
%
M(rms) = \left(5.75^{+0.39}_{-0.36}\right)\times10^7
\left(\frac{\lambda L_{\lambda}(5100\,{\mbox{\AA}} 
)}{\rm 10^{44}\,ergs\,s^{-1}}\right)^{0.402\pm0.034}M_{\odot}
\end{equation}
and is also plotted as a solid line.

The results of the above two methods are not consistent.  Moreover,
while it is arguable that using the rms spectra to determine $v$ is a
better method (see \S~\ref{secmass}), the mass--luminosity correlation
based on this measure is less significant. This can perhaps be
attributed to the fact that the line fluxes in the rms spectra are
weaker and hence the uncertainty in the corresponding FWHM larger.

Our M--L relation does not agree with the one found by Koratkar \&
Gaskell (1991b) of $M\propto L^{0.91\pm0.25}$\,, nor does it agree with
the one found by Wandel et~al. (1999) of $M\propto L^{0.77\pm0.07}$\,.
However, when Wandel et~al. (1999) use an unweighted linear fit they
find a slope of 0.54 which is in good agreement with our result.  We
have used a linear regression analysis which takes into account the
uncertainties in both coordinates (see \S~\ref{SML}).

The fact that the scatter in the mass--luminosity relation is larger
than that of the size--luminosity and velocity--luminosity relations
may indicate that luminosity, rather than mass, is the variable that
mainly determines the BLR size.  In fact, from the individual size and
velocity relations we expect:  $M\propto v^2R_{BLR} \propto
(L^{-0.27})^2 L^{0.7} = L^{0.16}$\,, i.e., a weak dependence of mass on
luminosity, with all AGN having similar masses. In practice, we have
found a somewhat stronger dependence, $M \propto L^{0.5\pm0.1}$\,, but
with a large scatter.

The M--L relation can be expressed in terms of the Eddington
luminosity, $L_{Edd}$.  Roughly estimating the bolometric luminosity as
$L_{bol}\approx 9\lambda L_{\lambda}$(5100\AA ), we obtain an Eddington
ratio of
\begin{equation}
\frac{L_{bol}}{L_{Edd}}\approx 0.13\left(\frac{\lambda
L_{\lambda}(5100\,{\mbox{\AA}} )}
{\rm 10^{44}\,ergs\,s^{-1}}\right)^{0.5}  \ \ .
\label{eqEddR}
\end{equation}
The Eddington limit, based on this rough estimate for $L_{bol}$, is
plotted as a dashed line in Fig.~\ref{fmvsl}. Some of the quasars in
our sample appear to be approximately at or beyond the Eddington
limit. According to Fig.~\ref{fmvsl}, the Eddington limit traces an
envelope in the mass--luminosity plane (if we ignore the two lower
right hand points, PG1700 and PG1704, which  have very large errors).
This depends, somewhat, on the factor chosen to estimate the bolometric
correction, i.e. the rather uncertain slope of the unobserved UV
continuum. However, if our chosen bolometric correction is realistic,
we may be seeing a direct indication that AGN energy is generated by
gas accretion.

Wandel (1999) reviews three classes of AGN mass estimation methods.
While the X-ray variability method (using the shortest time scale for
global luminosity variations as the light travel time across the
Schwarzschild radius) and the accretion disk modeling method (deriving
an accretion disk model that best fits the observed AGN continuum)
suggest that the Eddington ratio increases with luminosity, Wandel
(1999) notes that the kinematic methods (such as reverberation mapping)
have yet to show a similar trend. Our reverberation mapping result
(equation~\ref{eqEddR}) indicates for the first time that the Eddington
ratio increases with luminosity.

Models that suggest the bulk of the luminosity is due to energy release
via mechanisms which radiate up to a set fraction of the Eddington
luminosity are not consistent with the derived mass--luminosity
relation.  One such model is a geometrically thin, optically thick,
accretion disk that requires  ${L}/{L_{Edd}} < 0.3$ to be
self-consistent (Laor \& Netzer 1989). Our finding suggests that the
mass accretion rate grows with luminosity much faster than the central
mass, which would mean very different disk properties in low- and
high-luminosity sources. Our luminosity determination is based on the
monochromatic flux at 5100 \AA. In the thin disk model, this
monochromatic flux may represents a different fraction of the object's
bolometric luminosity in AGN of different masses. Thus, it is not clear
that the thin accretion disk model can be ruled out by the new
results.

\section{Summary}

Spectrophotometric monitoring of a large, optically-selected quasar
sample has shown clear correlations and well-defined time lags between
the optical continuum and the Balmer-line light curves. While the
Seyfert~1 galaxies that have been studied in this manner all have
optical luminosities $<$~1.5$\times$10$^{44}$~ergs~s$^{-1}$, the new
sample allows us to measure time lags in AGN with luminosities up to
$10^{46}$~ergs~s$^{-1}$. Our work increases the available luminosity
range for studying the size--mass--luminosity relations in AGN by two
orders of magnitude and doubles the number of objects suitable for
these studies. We have combined our results for 17 quasars with data
for 17 Seyfert~1 galaxies having reliable time lag measurements, and
derived uniform estimates of BLR size, central masses, and luminosities
for the combined sample.

Our main finding is that the BLR size scales with the
5100\,\AA\ luminosity as $L^{0.70\pm0.03}$. This is significantly
different from Wandel et al.'s (1999) analysis and is also in
contradiction with simple theoretical expectations, both suggesting
\Rblr$\propto L^{0.5}$. We have also found that the velocity field of
the BLR scales inversely with the luminosity, $v_{FWHM}\propto
L^{-0.27\pm0.01}$\,.  Combining the measured \Rblr\ with the observed
FWHMs, we have obtained a mass--luminosity relation for AGN, $M\propto
L^{0.5\pm0.1}$\,, which, however, has a large intrinsic scatter.  The
M--L correlations are based on two different estimates (mean and rms)
of the FWHM of the Balmer lines and are not consistent with each other,
despite the fact that the two measured values for the FWHM are
generally consistent. Empirically, at least, it is not obvious which
method of FWHM measurement is preferable.

Our results show the usefulness of long-term monitoring of
high-luminosity AGN. There is a need to expand the luminosity range to
include the highest luminosity quasars and this will require some 5--10
years of observations.  Follow-up studies are also needed for some of
the results obtained here, such as better determinations of the gas
distribution in the BLR and the exact SED of the quasars.  Future work
using this sample will include the study of time-variable line
profiles, lags between continuum bands, the intrinsic Baldwin relation,
and more.

\acknowledgments 

We are grateful to Brad Peterson for supplying us with the Seyfert~1
data and for many enlightening comments and discussions, and to the
referee, Martin Gaskell, for his constructive advice.  Ari Laor is
acknowledged for very useful discussions.  We thank John Dann and the
WO staff for their expert assistance with the observations throughout
the years.  Research at the WO is supported by grants from the Israel
Science Foundation. Monitoring of PG quasars at SO was supported by
NASA grant NAG 5-1630. H.N. \& S.K.  acknowledge financial support by
the  the Jake Adler Chair of Extragalactic Astronomy. S.K. acknowledges
financial support by the Colton Scholarships.

\end{document}